\documentclass[aps,prx,superscriptaddress,twocolumn,longbibliography]{revtex4-1}
\usepackage{bm} \usepackage{graphicx} \usepackage{amsmath}
\usepackage{amsthm,stmaryrd}
\usepackage{amssymb} \usepackage{epstopdf} \usepackage{txfonts}
\usepackage{color}

\newcommand{\ket}[1]{|{#1}\rangle}
\newcommand{\bra}[1]{\langle{#1}|}

\newcommand{\re}{\rm e}
\newcommand{\ri}{\rm i}

\DeclareMathOperator*{\Tr}{Tr}

\DeclareRobustCommand\openzero{\leavevmode\hbox{0\kern-.55em0}}
\mathchardef\minus="002D

\begin{document}

\title{
Driven quantum dynamics: will it blend?
}

\author{ Leonardo Banchi}
\affiliation{Department of Physics and Astronomy, University College London, Gower Street, WC1E 6BT London, United Kingdom}

\author{Daniel Burgarth }
\affiliation{ Institute of Mathematics, Physics, and Computer Science, Aberystwyth University, Aberystwyth SY23 2BZ, UK }

\author{Michael J. Kastoryano}
\affiliation{NBIA, Niels Bohr Institute, University of Copenhagen, Denmark}

\date{\today}
\begin{abstract}
Randomness is an essential tool
in many disciplines of modern sciences, such as cryptography, black hole physics, random matrix theory and Monte Carlo sampling. In quantum systems, random operations can be obtained via random circuits thanks to so-called $q$-designs, and play a central role in condensed matter physics and in the fast scrambling conjecture for black holes. Here we consider a more physically motivated way of generating random evolutions by exploiting the many-body dynamics of a quantum system driven with stochastic external pulses. We combine techniques  from quantum control, open quantum systems and exactly solvable models (via the Bethe-Ansatz) to generate Haar-uniform random operations in driven many-body systems. We show that any fully controllable system converges to a unitary $q$-design in the long-time limit. Moreover, we study the convergence time of a driven spin chain by mapping its random evolution into a semigroup with an integrable Liouvillean and finding its gap. Remarkably, we find via Bethe-Ansatz techniques that the gap is independent of $q$. We use  mean-field techniques to argue that this property may be typical for other controllable systems, although we explicitly construct counter-examples via symmetry breaking arguments to show that this is not always the case. Our findings open up new physical methods to transform classical randomness into quantum randomness, via a combination of quantum many-body dynamics and random driving.

\end{abstract}

\maketitle

\section{Introduction}

Randomness generating quantum operations play a central role in our  understanding 
of very various physical phenomena \cite{guhr1998random}. 
Recently, with the development of quantum information processing, 
random operations have found new applications,
not only as a theoretical tool,
but also in practical protocols. Indeed, they  
are used in 
quantum cryptography 
\cite{hayden2004randomizing},
quantum process tomography
\cite{bendersky2008selective},
fidelity estimation \cite{dankert2009exact},
quantum communication and entanglement sharing 
\cite{harrow2004superdense,abeyesinghe2009mother,hastings2009superadditivity}, 
quantum data-hiding 
\cite{divincenzo2002quantum,hayden2004randomizing,piani2014quantumness} 
and 
entanglement generation
\cite{oliveira2007generic,vznidarivc2008exact,hamma2012quantum,zanardi2014local}. 
Because of their crucial importance, several procedures have been developed 
to generate either truly random or pseudo-random operations 
via random quantum circuits 
\cite{brandao2016local,brown2010convergence,emerson2003pseudo,dankert2009exact,gross2007evenly,harrow2009efficient,turner2016derandomizing,alexander2016randomized}.
However, from the physical point of view, these protocols often have a complexity comparable  with universal quantum computation, being based on the 
application of a sufficiently large set of quantum gates. Here, on the other hand, 
we consider a more physically inspired approach, based on quantum control, 
where the quantum system is controlled 
by random classical  pulses.

Quantum control is an established research field at the overlap of control
theory and quantum mechanics. Essentially it provides a framework to steer a
quantum system through Hilbert space by applying time-dependent fields.
Controllability is a powerful algebraic tool to fully characterise when
{\it any} possible unitary evolution in the system's Hilbert space can be obtained 
from the Schr\"odinger equation with a suitable choice of time-dependent fields. 
The central question of this paper is what happens when we apply 
random fields to a controllable system. We will show, under some 
conditions, that after a suitably long {\it mixing } time 
the corresponding random unitary evolutions of the system converge 
to a uniformly random set, as measured by the Haar measure. 
Therefore, one of the central result of this paper is that driving a 
controllable quantum systems with stochastic control pulses offers
a natural approach to generate random unitary operations with physical 
processes.

Within this picture, the estimation of the mixing time is the crucial 
theoretical aspect. We use several tools from the theory of open 
quantum systems and many-body physics, such as low-energy effective
Liouvilleans, mean-field techniques and the Bethe-Ansatz, to find an accurate
estimation of the mixing time in several situations.  In particular, we 
focus on a one-dimensional system with edge control due to the availability
of analytical tools, as well as the intuitive interpretation available in such a
system with Lieb-Robinson bounds and spin waves. 
This particular case is also motivated by the 
current experimental capabilities in integrated photonic circuits
\cite{perez2013coherent,pitsios2016photonic}, where different stochastic 
control pulses can be simulated by changing the spatial extent of 
the waveguides via electrically tuned on-chip
heaters \cite{carolan2015universal}. In those systems a major recent result
has been the experimental measurement of {\it boson sampling}
\cite{broome2013photonic,spring2013boson,crespi2013integrated}, 
a problem which is believed to be hard to simulate classically. 
Random unitary operations and higher dimensional systems 
are required in boson sampling to have a 
convincing demonstration of 
quantum computational supremacy \cite{aaronson2011computational}. 
Pseudo-random operations in those experiments are currently obtained via 
a finely tuned network of several beam splitters and phase-shifters. 
The different approach presented here is based on the simpler implementation
of noisy quantum walks and, therefore, can offer an advantage to 
perform boson sampling experiments on larger scale. 

A further motivation for this paper comes from quantum control itself. 
The algebraic tools developed in quantum control are typically not
able to provide an estimation of the control time needed to reach a 
given target operation. 
In  view of practical applications, this
is a big handicap, because noise will always limit the total time available to
an experimenter. It is therefore of interest to find estimates of such times.
The analytical expressions for the mixing time obtained in this paper 
provide also an easily computable upper bound for the control time. 
Indeed, by definition, after the mixing time the system has already explored all 
possible unitary evolutions with stochastic control pulses. 
This implies that, apart from measure zero sets, at this time any evolution 
is achievable with a suitable choice of the control field. 

Finally, another motivation for the present work is 
for the problem of fast scrambling
of quantum information. The problem was first identified in the setting of
black hole physics \cite{hayden2007black,sekino2008fast}, 
where it was conjectured that black holes
start evaporating information when most localized microscopic degrees of
freedom become inaccessible without measuring a constant fraction of the whole
system. Unfortunately, identifying mechanisms for fast scrambling has been
challenging, and providing tools to rigorously analyze scrambling times even
more so. Moreover, 
explicit constructions of fast scramblers \cite{lashkari2013towards}
are not directly inspired by physical models. 
Here we describe a physically motivated process that
could lead to new insights in the design and analysis of fast
scrambling models.  

The paper is organized as follows: in section \ref{s:design} we show how 
to obtain Haar-uniform unitary evolutions (i.e. a unitary design) 
via quantum control techniques. We will focus on $q$-design, not only for its
applications in quantum information, but also to quantify the distance 
with the target uniform distribution. 
We 
will consider Markovian stochastic control pulses  and 
introduce some general techniques for the estimation of the mixing time. 
In section \ref{s:manybody} we map the problem of unitary design to a general 
many-body problem, studying its mean-field solution and discussing the 
limitations of the latter approach via symmetry breaking arguments. 
In section \ref{s:chain} we focus on a specific one-dimensional model
controlled at one of its boundaries. We show that this model in certain 
limits can be mapped to an exactly solvable model and we study its analytic
solution via Bethe-Ansatz techniques. 
A central result of this section is that 
the mixing time for this particular model is independent of the number of
copies $q$. 
Intuitively the $q$-independence implies that pseudo-random unitaries 
obtained with random control pulses approximate all the moments 
of the Haar distribution with the same accuracy. 
These predictions are then corroborated with numerical simulations. 
In Section \ref{s:appl} we show other applications 
for boson sampling, the decay of correlations in 
spin chains, and for the estimation of the control time. 
Conclusions and perspectives are written in section \ref{s:concl}.

\section{Unitary designs via quantum control}
\label{s:design}

Physical quantum systems are modeled via a Hamiltonian operator $H$, which
describes the interactions between the components of the system. 
When external control is applied to the system, its evolution is 
represented by a time-dependent Hamiltonian 
\begin{align}
  \hat H(t) = H + g(t) V~,
  \label{e:H}
\end{align}
where $g(t)$ is an external control pulse and $V$ is an operator. 
If $d$ is the dimension of the Hilbert space, then $H$ and $V$ are $d\times d$ 
Hermitian 
matrices while $g(t)$ is a scalar function depending on time $t$. 
For multiple pulses $\hat H(t)=H+\sum_i g_i(t) V_i$. After some time $T$,
the combined action of the natural interactions and the external pulses is 
a unitary operation 
$U= \mathcal T \exp\left({-}\ri \int_0^{T}\hat H(s)\, ds\right)$,
where $\mathcal T$ represents the time order operator. 
In general, the amount of different unitary operations $U$ that can be obtained from 
the dynamics of the system is limited. However, if the system is fully controllable,
then any operation can be obtained with a suitable engineering of the
control pulse. In other terms, 
given any $U\in{\rm SU}(d)$ it is possible to find a control profile
$g(t)$ such that 
$U= \mathcal T \exp\left({-}\ri \int_0^{T}\hat H(s)\, ds\right)$ 
where the control time $T$ depends on the target unitary
$U$. There are many powerful theorems to test controllability. In general a 
system described by the Hamiltonian as in Eq. \eqref{e:H} 
is controllable \cite{d2007introduction} if $H,  V$ 
and their nested commutators 
$[A, [B, [C, \dots]]]$ (where $\{A,B,C,\dots\}{\in}\{H,V\}$)
generate the Lie algebra of SU(d). 
Although the algebraic conditions for controllability are well known, 
it is still an open problem in quantum control to estimate the control time
$T$, given also the knowledge of the target gate $U$ and the operators 
$H$ and $V$. 
For fully controllable systems there exists a minimal control time, generally unknown,
such that all target gates can be obtained exactly at that time \cite{JV1972}. For small dimensional systems, analytic bounds of such 
\emph{universal control time} may be found
in terms of quantum speed limits or Cartan decompositions of spin systems. In
high dimensional system, such tools become intractable.  
If the system is drift-free ($H_0=0$), control times are trivial or only determined by
energy bounds on the time-dependent fields. We are instead interested in
systems where the controls need to work together with a drift to achieve full
control (so-called weak controllability). In such a case, the timescale is
bounded by the dynamics of the drift and provides insights into the many-body
physics triggered by it.

We now consider the control pulse as a stochastic process, 
namely where a certain 
profile $g(t)$ can be applied to the system
with a probability $p_{g(t)}$, 
and study the 
distribution of the resulting unitary operations. 
Such a random pulse can be obtained, for example, by considering 
the Fourier expansion of the control signal 
\begin{align}
  g(t) = \sum_{k=1}^K A_k \cos(\omega_k t + \varphi_k)~,
  \label{e:fourier}
\end{align}
where the amplitudes $A_k$, the phases $\varphi_k$, and possibly even the frequencies
$\omega_k$ are random variables. We use the notation $\mathbb E[\cdot]$ to denote the 
average over those random variables. Repeating the experiment with  
many random signals one obtains a distribution of unitary matrices, where each matrix $U$
is obtained with probability $p_U$. 
Random unitary operations play a central part in many quantum information
protocols. 
A pivotal role in many applications is played by the uniform 
distribution, called also Haar 
distribution, which is invariant under the action of the unitary group
itself. 
In the following sections we study when, and how rapidly, the distribution $p_U$ 
converges to the Haar-uniform distribution.

\subsection{Comparing random evolutions: unitary $q$-design }

Obtaining truly uniform random unitaries is a very hard 
task, and normally one observes  pseudo-uniform distributions which 
approximate the uniform (Haar) measure up to some errors. 
Pseudo-uniform distributions can be obtained with random quantum circuits
\cite{brandao2016local,brown2010convergence,emerson2003pseudo,dankert2009exact,gross2007evenly,harrow2009efficient},
but these circuits typically require many different gates that make 
demanding the implementation in physical systems. 
Recently, alternative protocols based on 
physically inspired time-dependent Hamiltonians have been proposed 
\cite{nakata2016efficient,onorati2016mixing}. Nonetheless, these 
approaches still require that all the interactions inside 
the system should change in time, an assumption that currently is beyond reach
in many experimental platforms.  Here, on the other hand, we focus on a general
scheme which occurs in most quantum systems, namely when the natural and
time-independent interaction $H$ experienced by the system  is paired with an
external control, as in Eq. \eqref{e:H}. 

There are many ways of comparing the distance between two quantum processes. When dealing with randomness generating processes, it is often convenient and relevant to work with approximate $q$-designs \cite{brandao2012local}. 
A unitary $q$-design is a distribution of unitaries, possibly discrete,
that gives the same expectations of the Haar distribution 
for polynomial functions of degree at most $q$
(see e.g. \cite{low2010pseudo}).  It is often inaccessible experimentally to distinguish between truely random processes and approximate $q$-designs. Formally, approximate $q$-designs are defined by the requirement that 

\begin{align}
  \left\| \mathbb{E}_{U} \left[U^{\otimes q} (\cdot) U^{\otimes q}{}^\dagger\right] -
  \int  U^{\otimes q} (\cdot) U^{\otimes q}{}^\dagger\,\mu_{\rm Haar}(dU)
\right\|_\diamond < \epsilon  ~,
  \label{e:design}
\end{align}
for suitably small $\epsilon$, where $\|\cdot\|_\diamond$ refers to the 
diamond norm, 
$\mathbb{E}_U$ denotes an average over some given
distribution of unitaries $\mu_U$ and $\mu_{\rm Haar}(dU)$ 
is the Haar measure. This is the most stringent distinguishability measure between quantum processes, and guarantees that no single (global) measurement on the system and a possible ancilla can distinguish between the two processes with probability larger than $\epsilon$.
A related notion \cite{harrow2009efficient}
is that of quantum expanders, which are defined by 
\begin{align}
  e(\mu_U, q)=\left\| \mathbb{E}_U\left[U^{\otimes q,q}\right] -
    \int  U^{\otimes q,q} \,\mu_{\rm Haar}(dU)\right\|_\infty < \epsilon  ~,
  \label{e:expander}
\end{align}
where $X^{\otimes q,q} = X^{\otimes q}\otimes (X^{\otimes q})^*$. 
Eq.~\eqref{e:expander} can be regarded as  the vertorised version of 
Eq.~\eqref{e:design}: given an 
operator $X=\sum_{ij}X_{ij}\ket i\bra j$, its {\it vectorized} form is 
$\vert X\rangle\rangle = \sum_{ij} X_{ij}\ket{ij}$. However it is striclty weaker, and the separation between the two bounds can be exponential in the system size. However, Eq.~\eqref{e:expander} is often much easier to work with in practice \cite{harrow2009efficient}. It follows from 
the definition that 
$\vert AX\rangle\rangle 
= A\otimes\openone \vert X\rangle\rangle$ and 
$\vert XA\rangle\rangle 
= \openone\otimes A^T \vert X\rangle\rangle$. Therefore, $X^{\otimes q,q}$ 
is the vectorization of the superoperator 
$\rho \mapsto X^{\otimes q} \rho X^{\otimes q}{}^\dagger$.  
Quantum expanders and $q$-design compare probability distributions of unitary matrices
by comparing the ``moments'' of the distribution, namely 
random processes that depend polynomially on the random variable. 
Two close distributions of unitary matrices have  similar {\it moments}, 
as shown in \cite{brandao2016local}, 
  $
  e(\mu, q) \le 2q \,\mathcal W(\mu, \mu_{\rm Haar})
  $,
for all measures $\mu$, being $\mathcal W$ the 
Wasserstein distance \cite{oliveira2009convergence}
$
  \mathcal W(\mu_1,\mu_2) = \sup_f \left|\int f(U)\,[\mu_1(dU)-\mu_2(dU)]~\right|
$,
where $f$ is a 1-Lipschitz function, and $U$ is a unitary matrix. The Wasserstein distance is a measure between classical probability distributions, and hence one can use a number of classical Markov chain mixing tricks to bound it. However, we will use not be using it, as we instead use tools from condensed matter physics to bound the mixing time.


In the quantum control setting, 
$\mathbb{E}_U$ in Eqs. \eqref{e:design} and 
\eqref{e:expander} is the average over many unitary operations 
obtained after the application of random pulses up to 
a certain time $T$. Therefore
\begin{align}
  \mathbb{E}_U \left[U^{\otimes q} \rho U^{\otimes q}{}^\dagger\right] 
   &= \mathbb{E}\left[\left(
\mathcal T {\re}^{{-}{\ri} \int_0^{T}\hat H(s)\, ds}\right)^{\otimes q}\,\rho\,\left(
\mathcal T {\re}^{{\ri} \int_0^{T}\hat H(s) \,ds}\right)^{\otimes q}\right]~.
\label{e:Et}
\end{align}
To simplify the theoretical description of this problem we 
make two assumptions. (i) We assume that  
the stochastic process $g(t)$ is Gaussian. 
This is a reasonable approximation in many-cases and can be obtained e.g. 
via Eq.~\eqref{e:fourier} when $K\gg1$,  
in view of the central limit theorem. (ii) We assume also that $g(t)$ is 
{\it harmonic}, namely that $\mathbb{E}[g(t+s)g(t)] = c(s)$ is independent of $t$. 
Moreover, without loss of generality, the harmonic process can be chosen
such that $\mathbb{E}[g(t)]=0$. 
In view of these assumptions, exploiting the results of 
\cite{ishizaki2009unified,banchi2013analytical}, 
in appendix~\ref{s:markov} we find 
a closed form expression for Eq.\eqref{e:Et}. 
That expression can be drastically simplified if we assume that 
the correlation time is finite and there exists a suitably large $T$ such that 
$T c(T s) \simeq \frac\sigma2 \delta(s)$ where $\delta$ 
is the Dirac delta function and
$\sigma$ is a constant. In the long-time limit, $t>T\gg\|H\|,\|V\|$, 
one finds then that 
\begin{align}
  \mathbb{E}_U \left[U^{\otimes q} \rho U^{\otimes q}{}^\dagger\right] 
  &\simeq  \re^{-t \mathcal L^q} \, \rho~,
\end{align}
where
\begin{align}
  \mathcal L^q \rho &= -\ri\left[H^{\oplus q}, \rho\right] - 
  \frac\sigma2\left[V^{\oplus q}, \left[V^{\oplus q}, \rho\right]\right]~, 
  \label{e:lindblad}
\end{align}
and $X^{\oplus q}{=}X{\oplus}X{\oplus}\dots$, being $\oplus$ the 
Kronecker sum $X{\oplus}Y{=}X{\otimes}\openone{+}\openone{\otimes}Y$. 
Therefore, with these three approximations, the long-time dynamics of the stochastic 
process is Markovian and described by the above Lindblad equation 
\cite{lindblad1976generators,gorini1976n}, where the operator $\mathcal L^q$ 
is called Liouvillean. 
Similarly to what happens with the replica trick in statistical physics 
\cite{mezard1987spin}, the average over the noise effectively 
couples the initially uncoupled copies. 
Sometimes we will use the more convenient vectorised form of the above equation 
\begin{align}
  {\mathcal L}_q = -\ri \mathring H^{\oplus q} -\frac{\sigma}2 
  (\mathring{V}^{\oplus q})^2
~,
  \label{e:Lvec}
\end{align}
where $\mathring X = X\otimes \openone-\openone\otimes X^T$ is the vectorization
of the commutator $[X,\cdot]$. 
If $t\to\infty$ then 
  $\mathbb{E}_U \left[U^{\otimes q} \rho U^{\otimes q}{}^\dagger\right] $
 converges to one of the  steady states
of the Liouvillean $\mathcal L^q$. 

In the following section 
we prove that the steady state manifold 
of $\mathcal L^q$ coincides with the state space after  
averaging over the Haar measure, namely that {\it all} the moments of the 
random unitary evolution converge to the averages over the uniform
distribution for $t\to\infty$. Moreover, we will study the mixing time via the 
gap of the Liouvillean and show that, in several cases, the latter 
is independent on $q$. Physically this is important, because it implies that 
all the moments converge 
(in 2-norm) at the same time, as given by the inverse of the Liouvillean gap,  
and that, accordingly, we can use the latter to estimate the mixing time of the 
random unitary evolutions. 

\subsection{Steady state of the Liouvillean evolution}
\label{s:steady}

We start by describing the steady state of $\mathcal L^q$. 
In general, the dimensionality of the steady state set is in one-to-one relation with
the conserved quantities of the Lindbladian evolution \cite{albert2014symmetries}. 
Given an orthonormal basis $\{M_\mu\}$  of the steady state space,
equipped with
the standard Hilbert-Schmidt product, 
there exists a dual operator set $\{J_\mu\}$ such that $
\mathcal L^{q\dagger} J_\mu=0$, where $\mathcal L^{q\dagger}$ is the Liouvillean operator 
\eqref{e:lindblad} after the substitution $H\to-H$. The latter substitution does not 
change the dynamical algebra, so algebraic considerations based on controllability hold
also for $\mathcal L^{q\dagger}$. 
From the conserved quantities $J_\mu$ and their dual operators $M_\mu$ one finds the 
steady state as
$\rho_\infty = \sum_\mu M_\mu \Tr(J_\mu \rho_0)$ where $\rho_0$ is the initial state
\cite{albert2014symmetries}.
Since the system is controllable, repeated commutators of 
$H^{\oplus q}$ and $V^{\oplus q}$ 
give rise to the algebra su$(d)^{\oplus q}$.  
Therefore, because of the Schur-Weyl duality 
\cite{goodman2000representations}, 
the only operators that commute with 
both $H^{\oplus q}$ and $V^{\oplus q}$, 
and more generally with 
Eq.~\eqref{e:Et}, are index permutation operators. 
Let $S_q$ be the group of permutations of the set $1,\dots,q$ and let 
$P_\sigma$, $\sigma\in S_q$ be 
the operator which permutes the index of the tensor copy 
$\mathcal H^{\otimes q}$, namely the operator that maps  
$  \psi_{i_1,i_2,\cdots,i_n} $ 
to $
  \psi_{\sigma(i_1),\sigma(i_2),\cdots,\sigma(i_n)}
$
for each set of indices $i_j$. 
It is simple to show that $P_{\pi}P_{\sigma}{=}P_{\pi\sigma}$ and that 
these operators form a unitary 
representation of the permutation group $S_q$. 
The index permutation operators are the only conserved quantities of the 
Liouvillean, $\mathcal L^q(P_\rho)=\mathcal L^{q\dagger}(P_\rho)=0$, so 
$\rho_\infty = \sum_\sigma \rho_\sigma P_\sigma$. However, since the 
operators $P_\sigma$ are not orthonormal, one has 
\begin{align}
  \Tr[P_\sigma^\dagger \rho_0] = \Tr[P_\sigma^\dagger \rho_\infty] = 
  \sum_{\pi\in S_q} \rho_\pi\Tr[P_\sigma^\dagger P_\pi]~,
\end{align}
where in the first equality holds because $P_\sigma$ is a conserved
quantity. 
By inverting the above equation we find that 
\begin{equation}
  \rho_\infty = \lim_{t\to\infty}e^{t \mathcal L_q}\rho_0 = 
  \sum_{\pi,\sigma} \left(M^{-1}\right)_{\pi\sigma} \; 
  \Tr\left[ P_\sigma^\dagger\,\rho_0\right] \; P_\pi, 
  \label{e:steadystate}
\end{equation}
where $M_{\sigma\pi}=\Tr[P_\sigma^\dagger P_\pi]$.
It has been shown in Ref. \cite{brandao2012convergence} 
that $M_{\sigma\pi}=d^{l(\sigma^{-1}\pi)}$ where $l(\sigma)$ is the number of 
cycles in the cycle decomposition of $\sigma$. The dimensionality of the 
steady state manifold is then given by the matrix rank of $M$. One finds 
that the steady state degeneracy is $\sim e^{\mathcal O(q)}$. 
The right-hand side of \eqref{e:steadystate} is exactly equal to the
integration over the Haar measure (see e.g. Proposition 3 in \cite{brandao2012convergence}). 
Therefore, we have shown that 
\begin{align}
  \lim_{t\to\infty}\re^{t \mathcal L^q}\rho = \int dU\, U^{\otimes q} \rho U^{\otimes q\;\dagger}
   ~, 
  \label{e:haarid}
\end{align}
namely that the infinite time-evolution of the system under the Liouvillean 
\eqref{e:lindblad} is equivalent to an integration over the Haar measure.

In summary,  we have shown that by driving 
a controllable system with random control pulses
Eq.~\eqref{e:fourier}, where the
stochastic process is Gaussian, harmonic and has a finite correlation time, 
then 
the resulting average evolution of the quantum system 
converges for $t\to\infty$ to a uniform integration
over the Haar measure.

\subsection{Construction of excited states}
\label{s:spectrum}

Certain excited states of the Liouvillean \eqref{e:Lvec} can be built up 
directly from the excitations of the individual quantum systems. 
It is convenient to separate $\mathcal L_q$ from Eq.~\eqref{e:Lvec} into local
terms ${\mathcal L}_k^{\rm loc}$ acting only on the $k$-th copy, and a non-local interaction. 
Indeed, 
\begin{align}
  \mathcal L_q &= \sum_{k=1}^q \mathcal L^{\rm loc}_k -\frac\sigma2\sum_{k\neq l=1}^q 
  \mathring{V}_k \mathring{V}_\ell~,
  \label{e:Ldecomp}
  \\
  \mathcal L^{\rm loc}_k &= -\ri \mathring H_k -\frac{\sigma}2 
  \mathring{V}_k^2~,
  \label{e:Lloc}
\end{align}
where 
$\mathring H_k$, $\mathring V_k$, and accordingly 
  $\mathcal L^{\rm loc}_k $, act only on the $k$-th copy. 
Therefore each  $\mathcal L^{\rm loc}_k $ for different $k$ is equivalent to a single-copy 
Liouvillean $\mathcal L_1$.  We assume that 
the operator $\mathcal L_1$ is diagonalizable (with right and left eigenvectors) and 
call 
\begin{align}
    \mathcal L_1 = \sum_i \lambda_i \Pi_{(i)}~,
    \label{e:L1diag}
\end{align}
its eivenvalue decomposition, where the eigenvalues $\lambda_j$ are ordered with decreasing
real part (starting from zero) and $\Pi_j$ are the corresponding eigenprojections. The operators
\begin{align}
    \Pi^{(i)}_{j}=
    \Pi^{\otimes (j-1)}_{(0)}\otimes\Pi_{(i)}
    \otimes\Pi_{(0)}^{(q-j)}~,
    \label{e:projq}
\end{align}
are then eigenprojections of $\mathcal L^q$, 
with eigenvalue $\lambda_i$. To show this, we 
 note indeed that $\Pi^{(i)}_{j}$ is proportional to the vectorization of the 
identity operator in each copy, aside from the $j$-th one, since
$\Pi_{(0)}$ is the projection onto the steady state and, accordingly, 
$\Pi_{(0)}(X)=\rho_\infty\Tr[X]$, which is proportional to the 
identity operator.  
Therefore, 
$\mathring{V}_l\,\Pi^{(i)}_{j}=0$ (because 
$[V_{l}, \Pi^{(i)}_{j}(X)]=0$ for all $X$), as
long as $l\neq j$. On the other hand, for $l=j$, it is  
$\mathring{V}_k\,\mathring{V}_l\,\Pi^{(i)}_{j} =0$, 
since by construction $k\neq j$. This shows that 
\eqref{e:projq} is a  projector on the eigenspace of $\mathcal L_q$ 
with eigenvalue $\lambda_i$. 
Moreover, from the operators \eqref{e:projq} one can also construct the eigenstates 
of $\mathcal L_q$ that act on the irreducible representations of the symmetric group
-- indeed since the permutation operators
$P_\sigma$ commute with the Liouvillean, then
$P_\sigma(\Pi^{(i)}_{j})P_\sigma^\dagger$ is an eigenprojection of 
$\mathcal L_q$ for all $\sigma$. 

In summary, the eigenstate of $\mathcal L_1$ with the lowest gap can be 
used to construct some exact eigenstates of $\mathcal L_q$, although it remains to be
shown that they have the smallest gap. These eigenvalues have degeneracy at least as large as the ground state degeneracy, since $P_\rho \Pi_j^{(i)}$ is also an eigenvector with eigenvalue $\lambda_j$ of $\mathcal{L}^q$.

\subsection{Convergence time} 
\label{s:convtime}

Given the results of the previous section, we want to know how rapidly the semigroup converges to the uniform distribution Eq. (\ref{e:haarid}). In Appendix \ref{s:smeiconv}, we provide a brief introduction to the convergence theory of dynamical semigroups, and argue that when the generator is not reversible (detailed balance), the convergence is governed by the singular value gap of the channels rather than the spectral gap of the generator. In general we want to bound the trace norm, but it will be more convenient to analyze the $2\rightarrow 2$ norm:
\begin{equation}||e^{t\mathcal{L}^q}-\mathcal U_\infty||_{1\rightarrow 1}\leq d^{2q}||e^{t\mathcal{L}^q}-\mathcal U_\infty||_{2\rightarrow 2},\end{equation}
where $\mathcal U_\infty = \lim_{t\to\infty}e^{t\mathcal L^q}$ and 
$d$ is the dimension of the local Hilbert space. 
Let $s_j(t)$ be the singular values of $e^{t\mathcal{L}}$, ordered from largest to smallest. The largest has magnitude one. Then the singular values of $(e^{t\mathcal{L}}-\mathcal U_\infty)$ are strictly smaller than one, and 
\begin{eqnarray}
||e^{t\mathcal{L}}-\mathcal U_\infty||_{2\rightarrow 2} &=& \sup_{\psi}|\langle \psi| (e^{t\hat{\mathcal{L}}}e^{t\hat{\mathcal{L}}^\dag}-\hat{\mathcal U}_\infty)|\psi\rangle|~.
\end{eqnarray}
If the Liouvillian were reversible, then the singular values $s_j(t)$ would be given by $e^{t\lambda_j}$, where $\lambda_j$ are the eigenvalues of $\mathcal{L}$. Unfortunately the semigroups that we will be working with are not Hermitian. Nonetheless, from
Eq. (\ref{eqn:appA}), we find that the $2\rightarrow 2$ norm can be bounded in terms of the eigenvalues and eigenvectors of $\mathcal{L}^q$ as 
\begin{equation}
||e^{t\mathcal{L}^q}-\mathcal U_\infty||_{2\rightarrow 2} \leq \sum_{j:\lambda_j\neq0} e^{t{\rm Re}[\lambda_j]}\sqrt{||R_j||~||L_j||},\label{eqn:evectsconv} \end{equation}
where $\lambda_j$ are the eigenvalues of $\mathcal{L}^q$, and $R_j,L_j$ are its right and left eigenvectors, satisfying ${\rm tr}[L^\dag_jR_k]=\delta_{jk}$. 

In general it is very difficult to bound Eq. (\ref{eqn:evectsconv}), since the norms of the eigenvectors can be very large, and it is often difficult to get good bounds on the spectrum. 
Nonetheless, in Appendices \ref{s:smeiconv}, \ref{s:weak} and \ref{s:strong}, we study both the weak $(\sigma=\epsilon\rightarrow 0)$ and strong 
$(\sigma=\epsilon^{-1}\rightarrow \infty)$ coupling limits, 
and show the following properties:
(i) the spectral gap is $O( \epsilon )$, both in the strong and weak 
coupling limits -- for strong driving, 
the decrease of the gap for larger
$\sigma$ is consistent with the general occurrence in open systems
\cite{zanardi2016dissipative}; 
(ii) the eigenvectors satisfy $|R_j\rangle =\mathcal{S}|\Phi_j\rangle$ and $|L_j\rangle = \mathcal{S}^{\dag,-1}|\Phi_j\rangle$, for some invertible matrix $\mathcal{S}$ and an orthonormal basis $|\Phi_j\rangle$. The condition number of $\mathcal{S}$ is $\kappa(\mathcal{S})\equiv ||\mathcal{S}||~||\mathcal{S}^{-1}||$ and satisfies  $\kappa(\mathcal{S})=O(1+\epsilon)$. 
Moreover, in sections \ref{s:mf} and \ref{s:chain} we will discuss some 
cases where the Liouvillean gap is independent on $q$. 
Models whose mixing time is independent on $q$ have been obtained also in 
\cite{onorati2016mixing}, at the expense of more stringent requirements 
on the fluctuating terms of the Hamiltonian. 

We then get that
\begin{equation}
||e^{t\mathcal{L}^q}-\mathcal U_\infty||_{2\rightarrow 2} \leq e^{-t\lambda^*}d^{2q}\kappa(\mathcal{S})^2,\end{equation}
where $\lambda^*$ is the eigenvalue with the smallest non-zero real part 
 and $\kappa(\mathcal S)=\mathcal O(1+\epsilon)$. 
In terms of the trace norm, we then get that 
\begin{equation}
\sup_\rho||e^{t\mathcal{L}^q}(\rho)-\mathcal U_\infty(\rho)||\leq e^{-t\lambda^*}d^{4q}\kappa(\mathcal{S})^2.\end{equation}
In the weak or strong coupling limits, the condition number will be of order one yielding a mixing time of $T^*\sim 4q\log(d)/\lambda^*$. We lost a lot in two steps of the bound, both times involving a term of order $d^{2q}$. In certain cases, this is overly pessimistic. For instances, for a tensor product of $n$ semigroups, the mixing time is $T^*\sim \log(n)T^*_1$, where $T^*_1$ is the mixing time of a single subsystem \cite{cutoff}. We might ask whether the mixing time of Eq. (\ref{e:lindblad}) is also of the order $T^*\sim \log(q)T^*_1$, with $T^*_1=O(1/\lambda^*)$? 

We can see that this is not the case from the following argument: 
\begin{eqnarray}
||e^{t\mathcal{L}^q}-\mathcal U_\infty||_{1\rightarrow 1}&\geq& ||e^{t\mathcal{L}^q}-\mathcal U_\infty||_{2\rightarrow 2}\\
&\geq&\sum_{j:{\rm Re}[\lambda_j]=-\lambda^*} e^{t{\rm Re}[\lambda_j]},
\end{eqnarray}
since the lower bound is saturated when $\mathcal{S}=\openone$, and we have isolated the subspace with eigenvalue $\lambda^*$. Now, in Section \ref{s:spectrum} we have argued that if the gap of $\mathcal{L}^q$ is the same as the gap of $\mathcal{L}^1$, then we can construct the eigenvectors with minimal non-zero eigenvalue of $\mathcal{L}^q$ from those of $\mathcal{L}^1$. In particular, the size of this subspace is at least as large as the size of the ground state subspace. But we know that the ground state subspace has dimension $d_0\geq e^{O(q)}$. Hence the first excited subspace does as well. Then,  
\begin{equation}
||e^{t\mathcal{L}^q}-\mathcal U_\infty||_{1\rightarrow 1}\geq e^{O(q)}e^{-t\lambda^*}.
\end{equation}
Thus the mixing time is at least $T^*\sim O(q/\lambda^*)$, even in the weak coupling limit.

Finally, we comment on the distinction between the singular value gap of $e^{t\mathcal{L}}$ and the eigenvalue gap of $\mathcal{L}$. We know that as $t\rightarrow \infty$, the singular value gap 
$s^*(t)$, namely the largest singular value $s_j(t)\neq 1$, converges to $e^{t\lambda^*}$, however it is not clear how rapidly this occurs. This will be discussed in the numerical studies 
of Sec. \ref{s:chain} where we will show that, both in the strong and weak coupling limits, 
the difference between the spectral gap and the 
singular value gap vanishes on a time scale much smaller than $1/\lambda^*$.

\section{Many-body theory of unitary design }
\label{s:manybody} 

In the previous section we have argued that bounding the spectral gap of the dynamical semigroup is in many relevant cases sufficient to get good estimates on the mixing time of the process. Here we will study such a gap 
by introducing a general mapping from a control Liouvillian to a 
non-Hermitian many-body Hamiltonian, and then study its mean field solution.   
The mean field approach has been already successfully 
applied \cite{brown2010convergence} to estimate the convergence time of 
permutationally invariant random quantum
circuits,  where at each step a gate from a universal set is
applied to a random pair of qubits. 
Moreover, in Sec. \ref{s:chain} we will analyze an integrable example via
Bethe-Ansatz techniques, 
from whose solution it appears that the eigenstates with smallest gap 
are constructed from the steady states by changing the internal state of a 
single unpaired particle. 
This fact shares several similarities with what happens in bosonic 
condensates, and in 
particular with their mean field solution \cite{blaizot1986quantum}. 
Motivated by these two examples, it is natural to apply the mean field analysis
to generic Hamiltonian evolutions with random pulses.  However, 
although the predictions of
the mean field solution are consistent with several numerical simulations, 
we will clarify
that   this approach cannot be general  by constructing  explicit
counterexamples via symmetry breaking arguments. 

\subsection{Mapping to a non-Hermitian many body Hamiltonian}
\label{s:hubbard}

A powerful method for estimating the spectral gap of the Liouvillean is to map 
Eq.~\eqref{e:Lvec} to a many-body problem, and then use powerful techniques developed in
condensed matter systems to obtain the spectrum. In order to find this mapping 
we introduce a basis $b_{\alpha\beta}=\ket\alpha\bra\beta$, $\alpha,\beta=1,\dots,d$
and call 
$B_{\alpha\beta}= b_{\alpha\beta}^{\oplus q}$. These operators satisfy the 
SU(d) commutation relation,
$[B_{\alpha\beta},B_{\gamma\delta}]=B_{\alpha\delta}\delta_{\beta\gamma} -
\delta_{\alpha\delta}B_{\beta\gamma}$ 
and therefore define a reducible representation of SU(d). 
Moreover, 
$
    \mathring X^{\oplus q} = X^{\oplus q}\otimes\openone - \openone\otimes X^{\oplus q}{}^T
     = \sum_{\alpha\beta} \left(
     X_{\alpha\beta} B^{\uparrow}_{\alpha\beta}  -
     (X^T)_{\alpha\beta} B^{\downarrow}_{\alpha\beta}\right)
$
where we set $B^{\uparrow}_{\alpha\beta} = B_{\alpha\beta}\otimes\openone$ and 
$B^{\downarrow}_{\alpha\beta} = \openone\otimes B_{\alpha\beta}$. 
Hence, the Liouvillean can be written as 
\begin{align}
    \mathcal L_q 
     &= -i \sum_{\alpha\beta} H_{\alpha\beta} (B_{\alpha\beta}^\uparrow -
    B_{\beta\alpha}^\downarrow) & 
    \label{e:liouv2sp}
    \\& \nonumber 
    \hspace{1cm} - \frac\sigma2 \sum_{\alpha\beta\gamma\delta} 
    V_{\alpha\beta} V_{\gamma\delta} ( 
    B^\uparrow_{\alpha\beta} - B^\downarrow_{\beta\alpha})(
    B^\uparrow_{\gamma\delta} - B^\downarrow_{\delta\gamma})
    ~.
\end{align}
The form \eqref{e:liouv2sp} is a convenient starting point because it depends only 
on the original $d\times d$ operators introduced in \eqref{e:H}, 
while the complicated action into the 
$q$-copy Hilbert space is transferred into the basis operators $B$. 

The operators $B$ form a reducible representation of SU(d) and can be decomposed 
in terms of irreducible operators that act on different invariant 
subspaces of the original $(\mathbb{C}^{d})^{\otimes q}$ Hilbert space. Indeed, because of
the Schur-Weyl duality, every
irreducible representation of $(\mathbb{C}^d)^{\otimes q}$ is decomposed as
$ (\mathbb{C}^d)^{\otimes q} = \otimes_\lambda \mathcal P^\lambda\otimes
U^{\lambda}$ where $\mathcal P^\lambda$ is an irreducible representation of 
the symmetric group $S_q$ and $\mathcal U^\lambda$ an irreducible representation of 
SU(d).  A convenient expression
for the fully-symmetric and fully-anti-symmetric subspaces is given by 
\cite{rowe2012dual}
$B_{\alpha\beta}=a_\alpha^\dagger a_{\beta}$, where $a_\alpha$ and $a_\alpha^\dagger$
are either bosonic or fermionic creation and annihilation operators. 
Moreover, even a generic (though reducible) representation can be constructed 
from either bosonic or fermionic annihilation operators by adding an extra index and 
writing $B_{\alpha\beta}= \sum_u a_{\alpha u}^\dagger a_{\beta u}$. From the definition
of $B$ one realizes that in this generic representation there are exactly $q$ particles since 
\begin{align}
  \sum_{\alpha u} a^\dagger_{\alpha u} a_{\alpha u} = 
  \sum_\alpha B_{ \alpha\alpha}= q \;\openone~.
  \label{e:numpart}
\end{align}
For convenience, we also perform the calculation in the
basis where $V$ is diagonal. Therefore, Eq.\eqref{e:liouv2sp} becomes
\begin{align}
    \mathcal L_q &= 
     -i \sum_{\alpha\beta u} H_{\alpha\beta} (a_{\alpha u\uparrow}^\dagger a_{\beta u\uparrow} 
    - a_{\beta u \downarrow}^\dagger a_{\alpha u \downarrow})  
    \\& \hspace{1cm} - \frac\sigma2 \sum_{\alpha\beta u v} 
    V_{\alpha\alpha} V_{\beta\beta} (n_{\alpha u\uparrow} - n_{\alpha u\downarrow})
     (n_{\beta v\uparrow} - n_{\beta v \downarrow})
    ~,
    \label{e:liouv2sym}
\end{align}
where $n_{x} = a_{x}^\dagger a_{x}$. Thanks to this general representation, the many-body 
Liouvillean has been mapped to a many-particle Hubbard-like problem \eqref{e:liouv2sym} 
where the hopping part is anti-Hermitian. The original dependence on $q$ is mapped to the 
number of particles, namely to the constraint \eqref{e:numpart} that there 
are exactly $q$ particles in the ``spin-up'' and ``spin-down'' states, 
$\sum_{\alpha u} n_{\alpha u \uparrow}
= \sum_{\alpha u} n_{\alpha u \downarrow} = q\,\openone$. 

\subsection{Mean-field approach} 
\label{s:mf}

We consider here the decomposition \eqref{e:Ldecomp} where
 each  $\mathcal L^{\rm loc}_k $ for different $k$ is 
 equivalent to a single-copy Liouvillean $\mathcal L_1$. 
From the above decomposition it is clear that if the gap of 
$\mathcal L^{\rm loc} = \sum_k \mathcal L^{\rm loc}_k $ equals the gap of $\mathcal L_q$
then the Liouvillean gap $\lambda^*$ is independent on $q$.

Extending the treatment of Section~\ref{s:hubbard},
we define a local basis of operators $\tilde{B}_{\tilde \alpha\tilde \beta} = 
B^{\uparrow}_{\alpha_\uparrow\beta_\uparrow} 
\delta_{\alpha_\downarrow\beta_\downarrow} 
+
\delta_{\alpha_\uparrow\beta_\uparrow} 
B^{\downarrow}_{\alpha_\downarrow\beta_\downarrow} 
$ 
where $\tilde \alpha=(\alpha_\uparrow,\alpha_\downarrow)$, and similarly for $\tilde \beta$,
are multi-indices running from 1 to $d^2$.  Therefore
we can write the decomposition Eq.~\eqref{e:Ldecomp} as 
\begin{align*}
    \mathcal L_q = &  \sum_{{\tilde{\alpha}},{\tilde{\beta}}} (\mathcal L_1^{\rm loc})_{{\tilde{\alpha}}{\tilde{\beta}}}
    \tilde{B}_{{\tilde{\alpha}}{\tilde{\beta}}}
    - \frac\sigma2
    \sum_{{\tilde{\alpha}},{\tilde{\beta}},{\tilde{\gamma}},{\tilde{\delta}}}
    \mathring V_{{\tilde{\alpha}}{\tilde{\beta}}} 
    \mathring V_{{\tilde{\gamma}}{\tilde{\delta}}} 
    \;
    (\tilde{B}_{{\tilde{\alpha}}{\tilde{\beta}}} \tilde{B}_{{\tilde{\gamma}}{\tilde{\delta}}}-
    \tilde{B}_{{\tilde{\alpha}}{\tilde{\delta}}}{\tilde{\delta}}_{{\tilde{\beta}}{\tilde{\gamma}}})~,
\end{align*}
and, writing $\tilde{B}_{{\tilde{\alpha}}{\tilde{\beta}}} = a_{\tilde{\alpha}}^\dagger a_{\tilde{\beta}}$ with bosonic 
operators, then 
\begin{align}
    \mathcal L_q & 
    = \sum_{{\tilde{\alpha}},{\tilde{\beta}}} (\mathcal L^{\rm loc}_1)_{{\tilde{\alpha}}{\tilde{\beta}}}\; a^\dagger_{\tilde{\alpha}} a_{\tilde{\beta}}
    - \frac\sigma2
    \sum_{{\tilde{\alpha}},{\tilde{\beta}},{\tilde{\gamma}},{\tilde{\delta}}}
    \mathring V_{{\tilde{\alpha}}{\tilde{\beta}}}\mathring V_{{\tilde{\gamma}}{\tilde{\delta}}}
    \;
    a^\dagger_{\tilde{\alpha}} a^\dagger_{\tilde{\gamma}} a_{\tilde{\beta}} a_{\tilde{\delta}}~.
    \label{e:lsb}
\end{align}
We assume that $\mathcal L_q^{\rm loc}$ is diagonalizable (with left and 
right eigenvectors) as 
$(\mathcal L_q^{\rm loc})_{{\tilde{\alpha}}{\tilde{\beta}}} =\sum_j 
Z_{{\tilde{\alpha}} j}\,\lambda_j\, Z^{-1}_{j{\tilde{\beta}}}$ for a non-singular matrix $Z$,
where $j=0$ corresponds to the steady state. Then
 we define new bosonic operators via the non-unitary Bogoliubov
transformation 
$ \tilde{a}'_i = \sum_{\tilde{\alpha}} Z_{{\tilde{\alpha}} i} a_{\tilde{\alpha}}^\dagger$,  
$ \tilde{a}_i = \sum_{\tilde{\alpha}} (Z^{-1})_{i{\tilde{\alpha}}} a_{\tilde{\alpha}}$. These operators still satisfy
the canonical commutation relations [$\tilde a_i,\tilde a_j']={\tilde{\delta}}_{ij}$, though
$\tilde a_i'\neq \tilde a_i^\dagger$. 
As shown in Appendix~\ref{s:emf}, in this language, the steady state of
the many-body Liouvillean 
\eqref{e:lsb} is therefore the 
boson ``condensate'' $\ket{\Omega}=\frac{(\tilde a_0')^q}{\sqrt q!}\ket 0$ where
$\ket 0$ is the bosonic vacuum. 
Elementary excitations with respect to this state can be constructed with 
a Bogoliubov (mean-field) approach by defining a variational 
wave-function $\ket{\psi} = \sum_j \psi_j 
\frac{(\tilde a_0')^{q-1}}{\sqrt{(q-1)!}}a_j\ket 0$, for $j\neq 0$ 
and optimising over the amplitudes $\psi_j$. These states are motivated by the 
analytic solution of the integrable model considered in Section \ref{s:chain}, 
where the excited states with minimal 
gap have a single quasi-particle excitation. 
Although 
mean-field techniques have been highly studied mostly for Hamiltonian systems 
\cite{blaizot1986quantum}, they can be extended also to non-normal 
operators \cite{laestadius2017analysis} where left and right eigenvectors 
form a bi-orthonormal basis. 
Within this variational formalism we show in Appendix~\ref{s:emf} that the
four-body interaction in \eqref{e:lsb} does not alter the eigenstates, which are 
therefore exactly given by the bare single-particle eigenstates 
$\ket{\Omega^{\rm exc.}_j}= \frac{(\tilde a_0')^{q-1}}{\sqrt{(q-1)!}}a_j\ket 0$
with exact eigenvalue $\lambda_j$, for any $q$. 
This shows that the eigenvalues, at least in the low-energy subspace, are not ``renormalized''
for larger values of $q$. 
The obtained states $\ket{\Omega^{\rm exc.}_j}$ are indeed the symmetric combination 
of \eqref{e:projq}, which,  as 
shown before, are an exact eigenstate of $\mathcal L_q$. Within this simple mean-field 
treatment there are no other eigenvalues with a smaller gap than 
$\min_j |\Re[\lambda_j]|$. 
Therefore, the final outcome of the mean field treatment is that, at least for fully
symmetric states, 
the Liouvillean gap is constant as a function of $q$.

\subsection{Counterexample to the mean-field treatment}

The mean field treatment of the previous section,
based on single particle excitations, predicts that the Liovillean gap
is independent on $q$, as long as the mean field 
approach  is accurate. 
Also the rigorous Bethe-Ansatz treatment of Section~\ref{s:chain}, 
valid for a particular integrable model, will show that the Liouvillean gap 
is independent on $q$, by explicitly showing that the states with minimal gap
are made by unpaired particles. That rigorous treatment thus justifies 
the mean-field approach, at least for that particular model. 
However, here we show that the predictions
of the mean-field theory cannot be general by finding a counterexample where a state with
two bounded particles (hence appearing for $q\ge2$) may have a lower gap. 

We construct this counterexample via symmetry arguments. 
Clearly in the fully controllable case $H$ and $V$ must not share a symmetry
-- otherwise only symmetric unitaries can be obtained -- but this lack of 
common symmetries is not sufficient. Indeed, generically, in tensor copies there 
may be other non-trivial symmetries but, because of the Schur-Weyl duality, in
the fully controllable case only the permutation symmetries can remain.
Suppose now that our system is not controllable because there exists
an operator $\tilde X$, different from a permutation operator, such that 
$[H^{\oplus p},\tilde X]=[V^{\oplus p},\tilde X]=0$ and that the solutions of 
$[H^{\oplus q},X]=[V^{\oplus q},X]=0$ for $q<p$ are only permutation operators. 
In this case, Eq.~\eqref{e:haarid} would be valid for $q<p$, but not when $p=q$,
as the symmetry $\tilde X$ introduces an extra steady state. 
Then, suppose that we restore full-controllability by adding a small 
$\mathcal O(\epsilon)$ term in either 
$H$ or $V$ such that the operator $\tilde X$ is not a symmetry anymore
(we say that the symmetry $\tilde X$ is explicitly broken). 
This splits the extra steady state into 
an eigenvector with small $\mathcal O(\epsilon)$ eigenvalue which, for small enough
$\epsilon$ can be smaller than the gap, obtained when $q<p$. If this counterexample 
can be constructed, then the gap for $q<p$ may be different from the gap at $q=p$. 
Below we show that this construction is indeed possible already with $p=2$ and 
that these extra eigenstates correspond to bound particles in the 
many-body framework.

As shown in Refs.~\cite{zeier2011symmetry,zimboras2015symmetry} a rather surprising 
necessary and sufficient condition for controllability is that there are 
exactly two independent solutions of the equations 
$[H^{\oplus 2},X]=[V^{\oplus 2},X]=0$. 
Nontheless, a simpler necessary condition 
(though not sufficient \cite{zeier2011symmetry}) is the absense of 
non-zero solutions to the set of equations 
\begin{align}
    QH^T + H Q = 
    QV^T + V Q = 0~.
    \label{e:tosh}
\end{align}
Taking the complex conjugate of Eq.\eqref{e:tosh} we find that 
$Q$ satisfies $ Q^*H + H^T Q^* = Q^*V + V^T Q^* = 0$, 
as $H$ and $V$ are Hermitian. 
Because of this, $QQ^*$ commutes with both $H$
and $V$ and, owing to the Schur's lemma, $Q Q^*$ is proportional to the identity. 
Refs.~\cite{obata1958subgroups} proved that $Q Q^*=1$ when $Q$ is symmetric and
$Q Q^*= -1$ when $Q$ is anti-symmetric. 
If there are non-zero solutions of \eqref{e:tosh}, then
the system is not controllable and there are extra steady states such as the
bosonic paired state for $q=2$
\begin{align}
    \ket{\psi_Q}= 
    \sum_{{\tilde{\alpha}}{\tilde{\beta}}} (Q\otimes Q^*)_{{\tilde{\alpha}}{\tilde{\beta}}}  a_{\tilde{\alpha}}^\dagger a_{\tilde{\beta}}^\dagger \ket 0~.
    \label{e:qss}
\end{align}
Indeed, for both $Q$ symmetric and anti-symmetric $Q\otimes
Q^*$ is symmetric, thus justifying the bosonic approach. 
The proof can be
readily obtained from \eqref{e:lsb}, indeed
for both $X=H,V$ 
\begin{align*}
    \mathring X_{{\tilde{\gamma}}{\tilde{\delta}}} a^\dagger_{\tilde{\gamma}} a_{\tilde{\delta}} \ket{\psi_Q} &= 
    \mathring X_{{\tilde{\gamma}}{\tilde{\delta}}} a^\dagger_{\tilde{\gamma}} a_{\tilde{\delta}} 
    \sum_{{\tilde{\alpha}}{\tilde{\beta}}} Q_{{\tilde{\alpha}}{\tilde{\beta}}}  a_{\tilde{\alpha}}^\dagger a_{\tilde{\beta}}^\dagger \ket 0
\\ &=
 [ 
    (X Q)\otimes Q^* + 
    (Q X^T)\otimes Q^* +\\&\phantom{=[} - 
    Q\otimes (X^T Q^*) - 
    Q\otimes (Q^* X) 
    ]_{{\tilde{\gamma}}{\tilde{\delta}}} a^\dagger_{\tilde{\gamma}}
    a_{\tilde{\delta}}^\dagger\ket0~,
\end{align*}
so because of Eq.\eqref{e:tosh} we find
$    \mathring H_{{\tilde{\gamma}}{\tilde{\delta}}} a^\dagger_{\tilde{\gamma}} a_{\tilde{\delta}} \ket{\psi_Q} 
=  \mathring V_{{\tilde{\gamma}}{\tilde{\delta}}} a^\dagger_{\tilde{\gamma}} a_{\tilde{\delta}} \ket{\psi_Q} =0$, namely 
$\mathcal L_2 \ket{\psi_Q}=0$. Hence,
the extra symmetry $Q$ introduces a pairing between bosons in the 
steady state, which is
expressed by Eq. \eqref{e:qss} -- note that it is indeed a pairing because 
$[Q\otimes Q^*]_{{\tilde{\alpha}}{\tilde{\beta}}} 
\neq Q_{\tilde{\alpha}}  Q^*_{\tilde{\beta}}$ since $Q$ is a matrix. 

As discussed before, we can restore controllability by explicitly breaking the
symmetry \eqref{e:tosh} with small terms:
$ QH^T + H Q = \epsilon_H $, 
$QV^T + V Q = \epsilon_V$ where 
at least one between $\epsilon_V$ or $\epsilon_H$ has to be 
non-zero, otherwise the system is not controllable. In this case $\ket{\psi_Q}$ is not 
a steady state but, within first order perturbation theory,  can be used to 
create a state with eigenvalue ${\tilde{\delta}}=\mathcal O(\epsilon_V,\epsilon_H)$. 
In particular, one can construct specific examples where 
$\epsilon_V$ and $\epsilon_H$ are much smaller than the gap $\lambda^*$ of $\mathcal L_1$
so that ${\tilde{\delta}}<\lambda^*$. Therefore, exploiting these broken symmetries
we can construct counterexamples where the gap changes as a function of $q$. 
The simplest example is a two spin system with 
$H=(\sigma_1^x\sigma_2^x + \sigma_1^y\sigma_2^y + \sigma_1^x) + \epsilon \sigma_1^z\sigma_2^z$
and $V=\sigma_1^y$, where $\sigma^{\alpha}_j$ are the Pauli matrices acting on the spin $j$. 
For instance, for $\epsilon=0.1$ the gap of $\mathcal L_1$ is $\approx 0.45$ while the gap of 
$\mathcal L_2$ is $\approx 0.05$. 

In spite of this counterexample, we have observed that in most numerical examples, 
performed for small values of $d$ and $q$ with a random choice of $H$ and $V$,
the Liouvillean gap is constant as a function of $q$. This allows us to conjecture 
that ``typically'', namely for most choices of $H$ and $V$, 
the Liouvillean has a constant gap, as predicted by the mean-field 
approach.  Since 
in Eq.~\eqref{e:Ldecomp} each copy interacts with all the others, 
this conjecture is supported by 
the well-known validity (see e.g. \cite{mezard1987spin})
of the mean-field solution in long-range models.

\section{The controllable quantum walk}
\label{s:chain}

\begin{figure}[t]
  \centering
  \includegraphics[width=0.35\textwidth]{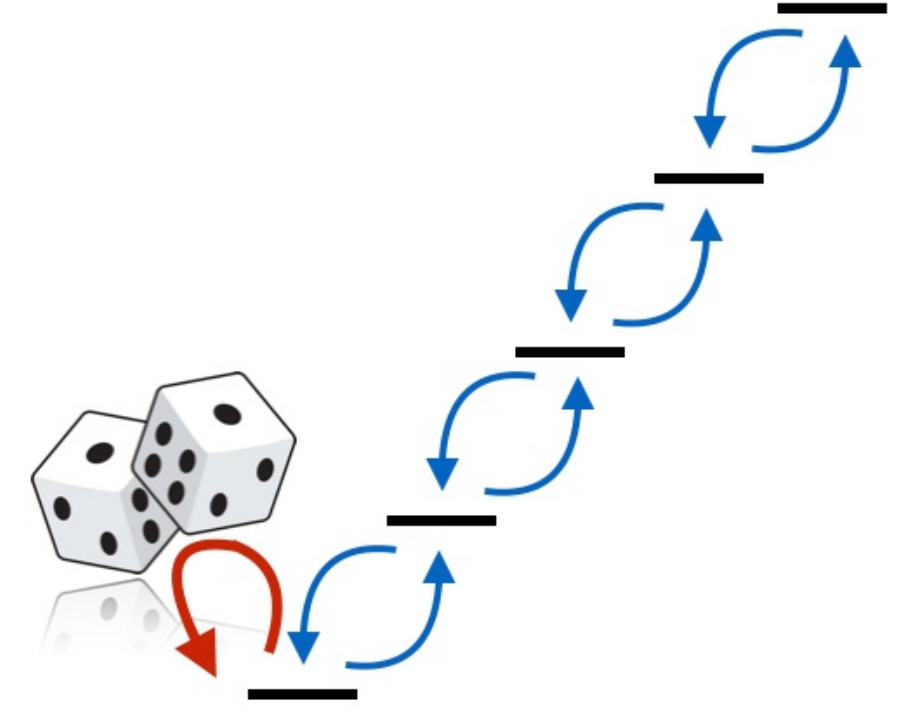}
  \caption{Depiction of a one dimensional quantum walk, with a local control at the bottom
    rung. Each site of the chain is coherently 
    coupled with its nearest neighbors. Random control pulses are applied to 
    the first site. 
  }
  \label{fig:cchain}
\end{figure}

We focus on a specific model that is of 
 experimental interest, namely a single-particle hopping 
in a one-dimensional lattice; see Fig. \ref{fig:cchain}. This framework can describe different
physical systems, such as a spin impurity in a spin chain, a single
electronic excitation in quantum dot arrays and a photon traveling in a 
one-dimensional photonic chip. 
The resulting {\it quantum walk} can be modeled via the Hamiltonian 
\begin{align}
  H = \sum_{n=1}^{L-1} \ket{n}\bra{n{+}1}+ {\rm h.c.}~,
  \label{e:H1}
\end{align}
where $\ket n$ represents the state in which the walker is in position
$n$, and $L$ is the length of the chain. This Hamiltonian 
has found numerous applications in quantum transport problems 
and remote entanglement generation in spin chains 
\cite{nikolopoulos2014quantum,banchi2011long,burgarth2005conclusive,banchi2011nonperturbative}. 

Moreover, we consider a local control field on a single site 
of the chain, namely the $c$-th site, which is modeled by Hamiltonian term
$g(t) V$, where
$V{=}\ket c \bra c$ and $g(t)$ is a time dependent control profile.
One can show that the chain is controllable provided that $c$ and $L{+}1$ 
are co-prime numbers \cite{wang2012symmetry,burgarth2013zero}. 
For simplicity, in the following we set $c=1$. 
The above hopping Hamiltonian with local control 
can be realized in many physical systems; for example, in reconfigurable
photonic chips \cite{perez2013coherent,pitsios2016photonic}, where the 
different control pulses can be obtained by electrically tuned on-chip
heaters \cite{carolan2015universal}.

In the following we evaluate the Liouvillean gap for all possible values of 
$q$ in the strong driving 
limit, namely when $\sigma\gg1$. The opposite weak driving limit is discussed 
in  appendix \ref{s:weak} for the single-particle $q=1$ case. 
We start by considering two important cases, namely the fully symmetric and 
fully anti-symmetric representation where $B_{\alpha\beta}=a^\dagger_\alpha a_\beta$ 
for either bosonic or fermionic degrees of freedom. We then extend our analysis 
to the general case.

\subsection{Gap analysis: fully-symmetric representation }
\label{s:symm}
We consider first the fully symmetric representation where 
$B_{\alpha\beta}=a^\dagger_\alpha a_\beta$ so one can omit the index $u$ 
from the equations of Section~\ref{s:hubbard}. Plugging the operators
$H$ and $V$ of the controllable chain into Eq. \eqref{e:liouv2sym} one 
finds the following 
Liouvillean 
\begin{align}
  \nonumber
    \mathcal L_q &= 
    -\ri \sum_{\alpha} (a_{\alpha\uparrow}^\dagger a_{\alpha{+}1,\uparrow} 
    - a_{\alpha\downarrow}^\dagger a_{\alpha{+}1,\downarrow} + {\rm h.c.})  
    \\& \hspace{1cm} - \frac\sigma2 
     (n^\uparrow_1 - n^\downarrow_1)
     (n^\uparrow_1 - n^\downarrow_1)
    ~.
    \label{e:liouv2chain}
\end{align}
To diagonalize the above operator we assume that $\sigma\gg1$ and we study the
``low-energy'' effective dynamics. In that limit the dissipative part $\sigma\mathcal
D=\frac\sigma2 
(n^\uparrow_1 - n^\downarrow_1) (n^\uparrow_1 - n^\downarrow_1)$ has either
eigenvalue 0 or $\sigma\gg1$. With a perturbative approach, 
discussed in Appendix~\ref{s:strong}, we decouple 
the latter ``high-energy'' subspace and obtain an effective Liouvillean
acting in the low-energy sector. From a first order expansion as a function of 
$\sigma^{-1}$  the effective Liouvillean is given by 
\begin{align}
    \hat{\mathcal L_q} = 
    & \frac2\sigma -\frac{2}{\sigma}\sum_{k=1}^{L-1} g_k  \Big[ -2(
        \tilde a_{0\uparrow}^\dagger \tilde a_{0\downarrow}^\dagger
        \tilde a_{k\uparrow} \tilde a_{k\downarrow} + {\rm h.c.}) + 
        \nonumber \\ & + 
        (\tilde n_{0\uparrow} + \tilde n_{0\downarrow} + 1)
    (\tilde n_{k\uparrow} + \tilde n_{k\downarrow} + 1) \Big] ~,
\label{e:Lstrongsym}
\end{align}
where 
$g_k = \frac{2}L
\sin^2\left(\frac{\pi k}L\right) $, 
$\tilde a_{k \updownarrow} = \sum_{\alpha=1}^{L-1}
\frac2L \sin^2\left(\frac{\pi k\alpha}L\right) a_{\alpha{+}1,\updownarrow}$ and 
$a_{1\updownarrow}\equiv \tilde a_{0\updownarrow}$. 
We call now 
$K_i^+ = \tilde a_{i\uparrow}^\dagger
\tilde a_{i\downarrow}^\dagger$, $K_i^- = (K_i^+)^\dagger$ and 
$K_i^z = (\tilde n_{i\uparrow} + \tilde n_{i\downarrow} + 1)/2$ and note that 
these operators satisfy the SU(1,1) commutation relations 
\begin{align}
    [K_i^+, K_i^-] &= -2 K^z_i~, & 
    [K_i^z, K_i^\pm] &= \pm K^\pm_i~,\\
    [K_i^\alpha,K_j^\beta] &= 0~, &  {\rm if ~} i&\neq j~.
\end{align}
With these definitions we find then 
\begin{align}
    \hat{\mathcal L_q} = 
    &\frac{2}\sigma  -
 \frac{8}{\sigma}\sum_{k=1}^{L-1} g_k  \, K_0 \cdot  K_k ~, 
 \label{e:gaudin}
\end{align}
where $ K_i \cdot  K_j \equiv -(K_i^+ K_j^- + K_i^- K_j^+)/2 + K_i^z K_j^z$
is the SU(1,1) invariant product,  namely the analogue of the Heisenberg
interaction. The model \eqref{e:gaudin} is a SU(1,1) Gaudin model
\cite{gaudin1976diagonalisation}, which is known to be exactly solvable 
with the Bethe-Ansatz approach. 
We explicitly diagonalize it in the appendix~\ref{s:richardson} by applying 
Richardson's method \cite{richardson1968exactly}. 
We find that the eigenvalues of the
Liouvillean $\hat{\mathcal L_q}$ are 
\begin{align}
\lambda = -\frac{2}\sigma \left(\sum_k g_k n_k   +
4\sum_{\alpha} E_\alpha\right)~,
\label{e:gaudinev}
\end{align}
where the non-negative integers $n_k$ parametrize the number of unpaired
particles in mode $k$ 
 (see the discussion in Appendix~\ref{s:richardson}) and the $E_\alpha$ are either zero or the solution of the non-linear set of equations
\begin{align}
\sum_k \frac{n_k + 1 }{\omega_\alpha -2 g_k^{-1}}  +
\frac{1}{\omega_k} + 2
\sum_{\beta\neq\alpha}\frac{1}{\omega_\alpha -\omega_\beta }
= 0~,
\label{e:gaudinomega}
\end{align}
where
$E_\alpha=1/\omega_\alpha$.
From that expression it is clear that the steady state
corresponds to $E_\alpha=0$ and $n_k=0$, for each $\alpha$ and $k$. 
Solutions to the above equations are known to be related with the roots 
of Heine-Stieltjes polynomials (see e.g. \cite{szego1959orthogonal}). 
By exploiting this relationship, one finds that all the solutions 
$\omega_\alpha$ of \eqref{e:gaudinomega} are real, 
different from each other, and different from the poles of
\eqref{e:gaudinomega}. Moreover, $g_k = g_{L-k}$ so the sum in
\eqref{e:gaudinomega} can be restricted to the first half where $g_k < g_{k+1}$. 
The roots of the Heine-Stieltjes polynomials have also the important property  that they
lie inside the intervals $2g_{k+1}^{-1} < \omega_\alpha < 2 g_k^{-1}$ for some $k$, 
so that  $2 E_\alpha > \min_k g_k = g_1$.
This constraint allows us to find rigorously the gap of the Liouvillean 
$\hat{\mathcal L_q}$. Indeed, thanks to the latter inequality, 
the paired states have a larger gap than the unpaired ones,
so we can focus only on  the solutions where $E_\alpha=0$. The minimum gap is
then obtained when $n_1 = n_{L-1}=1$ and $n_k=0$ otherwise. This is an allowed
state (for $L>2$) as it satisfies all the constraints and provides the gap 
\begin{align}
    {\rm gap} \equiv \lambda^* = \frac{8}{\sigma L } \sin^2\left(\frac \pi L\right)
    = \mathcal O(L^{-3})~.
    \label{e:gapfinal}
\end{align}
This gap is exact in the strong driving limit, can be achieved already at $q=1$ and is the same for all higher values
of $q$, as we have shown that there are no smaller non-zero eigenvalues. 
Therefore, we proved  here explicitly that in the strong driving limit the gap is independent on the number of copies $q$. 
In the following sections we extend this result, which up to now is restricted 
to the fully-symmetric representation, to show that 
\eqref{e:gapfinal} is indeed the gap, 
irrespective of the chosen representation.

\subsection{Gap analysis: anti-symmetric representations}
We first consider another particular case,
namely the fully anti-symmetric representation, that will
be used as a basis for the general solution discussed in 
the next section. 
We start from \eqref{e:liouv2sym} and we write $B^\alpha_{ij} = a_{i\alpha}^\dagger
a_{j\alpha}$ with fermionic creation and annihilation operators. 
Repeating the effective Liouvillean description of the previous section we find
\begin{align}
    \hat{\mathcal L_q} 
        &=-\frac{2}\sigma  +
    \frac{8}{\sigma}\sum_{k=1}^{L-1} g_k  \, S_0 \cdot  S_k ~, 
    \label{e:gaudinS}
\end{align}
where $ S_0 \cdot  S_k = \sum_{\alpha=x,y,z} S_0^\alpha S_k^\alpha$ refers to the SU(2)-invariant, product, namely the 
spin Heisenberg  interaction,  $S^\pm_j = S_j^x\pm i S_j^y$, and where we have defined   
$S_j^- = \tilde{a}_{j\uparrow} \tilde{a}_{j\downarrow}$,
$S_j^+ = (S_j^-)^\dagger$ and 
$S_j^z = (\tilde{a}_{j\uparrow}^\dagger \tilde{a}_{j\uparrow} + 
\tilde{a}_{j\downarrow}^\dagger \tilde{a}_{j\downarrow} -1)/2$. It is simple to verify 
that the above operators satisfy the SU(2) commutation relations on the same site, and commute on different sites, so that Eq. 
\eqref{e:gaudinS} is equivalent to the central spin model first studied by Gaudin 
\cite{gaudin1976diagonalisation}.  The diagonalization of the Gaudin
Heisenberg Hamiltonian proceeds along the same lines of the SU(1,1) one. 
There are two main differences: (i) the different sign in \eqref{e:gaudinS} and
\eqref{e:gaudin} and (ii) becayse of the Pauli exclusion principle the number of particles $n_k$ per mode $k$ is limited to either 0 or 1.  We find then that the eigenvalues
are given by Eq.~\eqref{e:gaudinev}, where the non-zero energies
$E_\alpha$ are the solutions of
\begin{align}
    \sum_k \frac{g_k (n_k -1)\ }{2E_\alpha -g_k} - 2
    \sum_{\beta\neq\alpha}\frac{E_\beta}{E_\alpha - E_\beta}
= 1~.
\label{e:gaudineqNZf}
\end{align}
However, because of the different sign in \eqref{e:gaudineqNZf}, we 
cannot relate the solutions of \eqref{e:gaudineqNZf} to the roots 
of the Heine-Stieltjes polynomials, so we cannot bound the gap using 
the argument of the fully symmetric case. 
Nonetheless, in the next section we consider a more 
general technique, valid for all the representations, where such a 
bound can be obtained using physical arguments borrowed from 
classical electrostatics.

\subsection{General gap analysis}
\label{s:gapgen}
As we have discussed in Section~\ref{s:hubbard}, 
a general representation of the SU(L) algebra can  can be obtained via extended creation and annihilation
operators \cite{rowe2012dual}, namely $B_{\alpha\beta}=\sum_u \tilde{a}^\dagger_{\alpha u } \tilde{a}_{\beta u}$ for either bosonic 
or fermionic operators. 
We use the fermionic representation for convenience, 
since our derivation uses the particle-hole symmetry that is a non-unitary operation 
in bosonic systems (see e.g. \cite{blaizot1986quantum}). Because
of the Pauli exclusion principle, in order to satisfy the constrain
$\sum_\alpha B_{\alpha\alpha} =q$, the auxiliary index $u$ has to run 
from 1 to $q$. Performing the same perturbative approach of Appendix~\ref{s:strong}, valid in the strong driving limit $\sigma\gg 1$, one finds that 
the effective Liouvillean $\hat{\mathcal L_q}$ can be written in the diagonal basis of 
the Hamiltonian as 
\begin{align}
  \nonumber
  \hat{\mathcal L_q} = -\frac2\sigma\sum_{k=1}^{L-1} g_k\Big(&
    B^\uparrow_{0k}B^\uparrow_{k0} + 
    B^\uparrow_{k0}B^\uparrow_{0k} + 
    B^\downarrow_{0k}B^\downarrow_{k0} + 
    B^\downarrow_{k0}B^\downarrow_{0k}
    \\ &- 2 B^\uparrow_{0k} B^{\downarrow}_{0k}
    - 2 B^\uparrow_{k0} B^{\downarrow}_{k0}
  \Big). 
  \label{e:gaudingen}
\end{align}
The above Hermitian operator corresponds to the purely dissipative Liouvillean 
\begin{align}
  \nonumber
  \hat{\mathcal L^q}\rho = -\frac2\sigma\sum_{k=1}^{L-1} g_k \left(
  \left[\tilde{V}_k^{\oplus q}, \left[\tilde{V}^\dagger_k{}^{\oplus q}, \rho\right]\right]
  + \left[\tilde{V}^\dagger{}_k^{\oplus q}, \left[\tilde{V}_k^{\oplus q}, \rho\right]\right]
  \right)~,
\end{align}
where $\tilde{V}_k = \ket{\omega_k}\bra{\omega_0}$, being 
$\ket{\omega_k}=\sum_{j=1}^{L-1}\frac2L\sin\left(\frac{\pi jk}{L}\right)^2\ket{j{+}1}$ and 
$\ket{\omega_0}=\ket{1}$.
One can check that the operators $\tilde{V}_k$ and their 
Hermitian conjugate form a controllable set, so the steady state of the effective
Liouvillean coincides with the original one. 
We now perform two transformations. The first one is the Jordan-Wigner 
transformation to obtain proper fermionic degrees of freedom, namely
where creation/annihilation operators with different indices $\uparrow$ 
and $\downarrow$ anti-commute. The second-one is a particle-hole transformation
in the spin-down sector. These transformations are implemented together by
  defining 
$W=\prod_{ju} e^{i \tilde{a}^\dagger_{ju\uparrow} \tilde{a}_{ju\uparrow}}$ 
and setting $a_{ju\uparrow} = \tilde{a}_{ju\uparrow}$ and 
$a_{ju\downarrow} = W\tilde{a}^\dagger_{ju\uparrow}$. 
Eq. \eqref{e:gaudingen} then becomes 
\begin{align}
  \hat{\mathcal L_q} &= -\frac2\sigma\sum_{k=1}^{L-1} g_k \sum_{\alpha\beta}
  \left[
    a^\dagger_{0\alpha} a_{k\alpha}
    a^\dagger_{k\beta} a_{0\beta} + 
    a^\dagger_{k\alpha} a_{0\alpha}
    a^\dagger_{0\beta} a_{k\beta}\right]
  \nonumber
  \\
   &= -\frac{2q}\sigma+\frac4\sigma\sum_{k=1}^{L-1} g_k 
   \sum_{\alpha\beta}
   X^{(0)}_{\alpha\beta}
   X^{(k)}_{\beta\alpha}~,
   \label{e:gaudingl}
\end{align}
where 
$
X^{(j)}_{\alpha\beta}= \left(a^\dagger_{j\alpha} a_{j\beta} - 
	a_{j\beta} a^\dagger_{j\alpha}\right)/2 
$
and the Greek letters refer to the  multi-index composed by the auxiliary index and the ``effective spin'' index, i.e. $\alpha=(u s)$  where $u=1,\dots,q$ and
$s=\{\uparrow,\downarrow\}$.  
The  traceless 
operators $X^{(j)}_{\alpha\beta}$ 
satisfy the SU(2q)$^{\oplus L}$ commutation relations, 
\begin{equation}
    [X^{(j)}_{\alpha\beta}, X^{(k)}_{\gamma\delta}] = \delta_{jk}
   \left(X^{(j)}_{\alpha\delta}\delta_{\beta\gamma} - X^{(j)}_{\gamma\beta}
   \delta_{\alpha\delta}\right), 
 \label{e:sualgebra}
\end{equation}
so that Eq.\eqref{e:gaudingl} represents a SU(2q) version of the Gaudin model. 
Indeed, Eq.\eqref{e:gaudingl} is invariant under the Bogoliubov
transofmation $a_{j\alpha}\to\sum_\beta U_{\alpha,\beta} \,a_{j\beta}$,
where $U$ is a unitary $(2q)\times(2q)$ matrix. 
SU(2q) has $(2q)^2-1$ generators, so one operator in 
\eqref{e:sualgebra} is dependent on the others. This is shown by
the equation $[\sum_\alpha X^{(j)}_{\alpha\alpha}, X^{(k)}_{\beta\gamma}]=0$ 
for each $\beta$ and $\gamma$. 
Going back to the original representation, namely performing back the 
particle-hole transformation, one finds that 
\begin{subequations}
\begin{align}
    X^{(j)}_{(x,\uparrow),(y,\uparrow)} & = 
    \frac{\tilde{a}^\dagger_{jx\uparrow} \tilde{a}_{jy\uparrow} - 
      \tilde{a}_{jy\uparrow} \tilde{a}^\dagger_{jx\uparrow}}2 ~,
    \\
    X^{(j)}_{(x,\downarrow),(y,\downarrow)} & = 
    \frac{\tilde{a}_{jx\downarrow} \tilde{a}^\dagger_{jy\downarrow} - 
      \tilde{a}^\dagger_{jy\downarrow} \tilde{a}_{jx\downarrow}}2 ~,
    \\
    X^{(j)}_{(x,\uparrow),(y,\downarrow)} & = 
    \tilde{a}^\dagger_{jx\uparrow} W\tilde{a}^\dagger_{jy\downarrow}~,
    \\
    X^{(j)}_{(x,\downarrow),(y,\uparrow)} & = 
    \tilde{a}_{jx\downarrow} W\tilde{a}_{jy\uparrow}~.
\end{align}
\label{e:su2q}
\end{subequations}
The Gaudin-like model \eqref{e:gaudingl} has been solved for different 
algebras (namely not only the SU(1,1) and SU(2) cases discussed before) 
in Refs.~\cite{ushveridze1994quasi,falceto1997unitarity}, while the duality between the different models that 
can be obtained by exploiting the auxiliary indices has different ramifications in mathematical physics (see e.g. \cite{mukhin2008bispectral} and references therein), especially due to its connections with the Knizhnik-Zamolodchikov equation \cite{mukhin2008bispectral,feigin1994gaudin}. 
In Appendix~\ref{s:gaudinsuq} we exploit the general 
solution \cite{ushveridze1994quasi,falceto1997unitarity} of the Gaudin model \eqref{e:gaudingl}, valid 
when the operators $X$ define any semi-simple Lie algebra, 
to obtain  the eigenvalues of the Liouvillean \eqref{e:gaudingl} when the SU(2q) operators are defined via the fermionic representation \eqref{e:su2q}. 
As in the fully-symmetric and fully-antisymmetric case 
discussed in the previous sections, the eigenvalues of 
$\hat{\mathcal L}_q$ are parametrized by non-negative
integers $n_{\uparrow j}$ and $n_{\downarrow j}$, 
and are given by 
\begin{align}
\lambda  = -\frac2\sigma\left[
\sum_{k=1}^{L-1}  g_k\left(
n_{\downarrow k} + n_{\uparrow k}   \right)
+ 4 \sum_\alpha \frac{1}{\omega_{q,\alpha} } \right]~,
\label{e:gaudinevgen}
\end{align}
where 
$\omega_{j,\alpha}$ for $j=1,\dots,2q-1$ are the solutions of 
\begin{align}
\sum_{\beta} \frac{2}{\omega_{j,\beta}-\omega_{j,\alpha}} =& 
\sum_{k=0}^{L-1} \frac{\mu^k_j}{z_k- \omega_{j,\alpha}} + 
\label{e:bethesuq}
\\&+
\nonumber
\sum_{\beta} \frac{1}{\omega_{j+1,\beta}-\omega_{j,\alpha}} +
\sum_{\beta} \frac{1}{\omega_{j-1,\beta}-\omega_{j,\alpha}} 
~,
\end{align}
being $z_0=0$, $\mu^0_j=\delta_{qj}$, and, for $k>0$, $z_k=2g_k^{-1}$ and
$
\mu^k_j = \delta_{j,q}(1-\delta_{n_{\downarrow k}>0} - \delta_{n_{\uparrow k}>0})
+ \delta_{j,q+n_{\uparrow k}} + \delta_{j,q-n_{\downarrow k}}
$. In \eqref{e:bethesuq} we set $\omega_{0,\beta}=\omega_{2q,\beta}\to-\infty$ namely, in other 
terms, for $j=1$ or $j=2q-1$ one of the two fractions in the second line is zero.

Owing to the similarity between Eqs.~\eqref{e:gaudinevgen} and \eqref{e:gaudinev}, if we can show that the solutions of \eqref{e:bethesuq}
satisfy the inequality  $2\omega^{-1}_{q,\alpha}> g_k$ for each $\alpha$ and $k$, then we can straightforwardly apply the reasoning of
Section~\ref{s:symm} to prove that the gap is indeed given by 
Eq.~\eqref{e:gapfinal} for any representation.  
However, the sign difference between Eqs.~\eqref{e:bethesuq} and 
\eqref{e:gaudinomega} prevents us from  using the theory of Heine-Stieltjes polynomials to prove that inequality, as we did in Section~\ref{s:symm}. 
Here we use a different approach, used also in Ref.~\cite{ushveridze1994quasi} for a different purpose, which is based 
on mapping the mathematical equations \eqref{e:bethesuq} to an electrostatic problem, and then use our classical physics intuition. Following Ref.~\cite{ushveridze1994quasi} we define the two-dimensional vector 
$\vec{\omega}_{j\alpha}$ whose real components are the real  and imaginary
part of $\omega_{j\alpha}$ and  interpret those vectors as 
the positions of some  particles with index $\alpha$ and  species  
$j=1,\dots,2q-1$. The equations \eqref{e:bethesuq} can then be interpreted 
as the conditions for an extremum of the function $\mathcal W(\{\omega\})$ defined as 
\begin{align}
\mathcal W(\{\omega\}) &= -\sum_{i,j=1}^{2q-1} \sum_{\alpha\beta} C_{ij} 
\log\left|\vec{\omega}_{i\alpha}-\vec{\omega}_{j\beta}\right| - \sum_{i=1}^{2q-1} \sum_{\alpha} \mathcal V_i(\vec{\omega}_{i\alpha}) ~,
\label{e:coulomb} \\
\mathcal V_i(\vec{\omega}) &= -\sum_{k=0}^{L-1} \mu^k_i \log\left|\vec{\omega}-\vec{z}_k\right|~,
\label{e:pot}
\end{align}
where $\vec{z}_k = (z_k,0)$ and the Cartan matrix $C_{ij}$ 
has non-zero components only on the diagonal, 
where $C_{ii}=2$, and for $|i-j|=1$, where $C_{ij}=-1$. 
\begin{figure}[t]
  \centering
  \includegraphics[width=.4\textwidth]{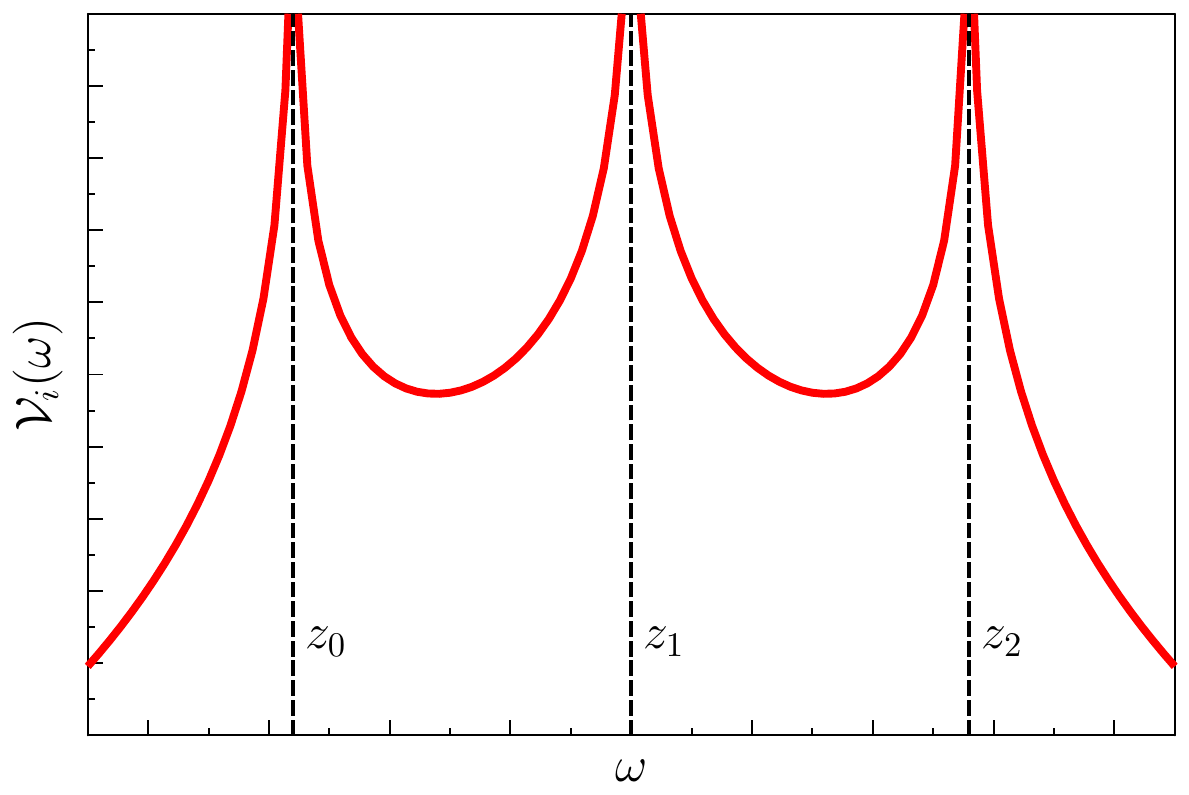}
  \caption{Example one-dimensional potential $\mathcal V_i(\omega)$ from Eq.~\eqref{e:pot}
  with three different values of $z_k$ and $\mu^k_i=1$.}
  \label{fig:pot}
\end{figure}
This shows that the problem of finding a solution to the system of equations
\eqref{e:bethesuq} is equivalent to the problem of finding the 
equilibrium positions of a set of particles in a two-dimensional plane 
interacting via the logarithmic potential \eqref{e:coulomb}. That potential is
analogous to the electrostatic potential since the Coulomb interaction in 2D is
logarithmic. Particles of the same species repel each other, while particles
with nearest-neighbour species attract each other. Finding the equilibrium
positions of those particles is in general quite complicated, although the
problem can be solved explicitly in the thermodynamic limit
\cite{roman2002large}. At first sight one may think that the problem has no
solutions since the potential \eqref{e:pot} is unstable. However, because of
the Z$_2$ symmetry ($\Im[\omega_{j,\alpha}]\to-\Im[\omega_{j,\alpha}]$), due to
the fact that the  $z_k$'s are reals, all the forces on the real line are
longitudinal. This property allows us to seek for
solutions of Eq.~\eqref{e:bethesuq} in the class of real numbers 
\cite{ushveridze1994quasi}. On the real line, the problem becomes stable and one-dimensional. 
An example of this effective one-dimensional potential is shown on Fig.~\ref{fig:pot}
where one can see the two unbounded regions for $\omega<\min_k z_k$ and for
$\omega>\max_k z_k$, where no solutions can exist.  
Therefore, this electrostatic analogy shows that the only stable solutions with finite 
$\omega_{i\alpha}$ can be found only between poles of $\mathcal V_i(\omega)$, or, 
in other terms, that the solutions of the non-linear set of equations \eqref{e:bethesuq} 
satisfy the constraint $\min_k z_k< \omega_{j\alpha} < \max_k z_k$, i.e. 
$2\omega_{j\alpha}^{-1}> \min_k g_k$. This, together with the discussion of 
Section~\ref{s:symm}, shows that  Eq.\eqref{e:gapfinal} is indeed 
the gap of the Liouvillean 
$\hat{\mathcal L}_q$ in the strong-driving limit.

\subsection{Numerical results for the controllable chain}
\label{s:numerics}

In the previous sections 
we have done an extensive theoretical analysis to show 
that, in a chain controlled on one boundary, the Liouvillean gap in the 
strong-driving limit is constant 
as a function of $q$ and scales as $\propto L^{-3}$ as a function of the 
length $L$ of the chain -- this scaling is consistent with what has been obtained in 
spin chains with boundary dissipation \cite{vznidarivc2015relaxation}. 
The scaling $\propto L^{-3}$ is obtained also in the
weak driving limit discussed in Appendix~\ref{s:weak}, though that analysis is 
valid only for $q=1$. Nontheless, in all our numerical experiments obtained 
for small values of $L$ and $q$ we found that the gap is constant as a function of 
$q$ over the whole range of $\sigma$. 
\begin{figure}[t]
  \centering
  \includegraphics[width=0.45\textwidth]{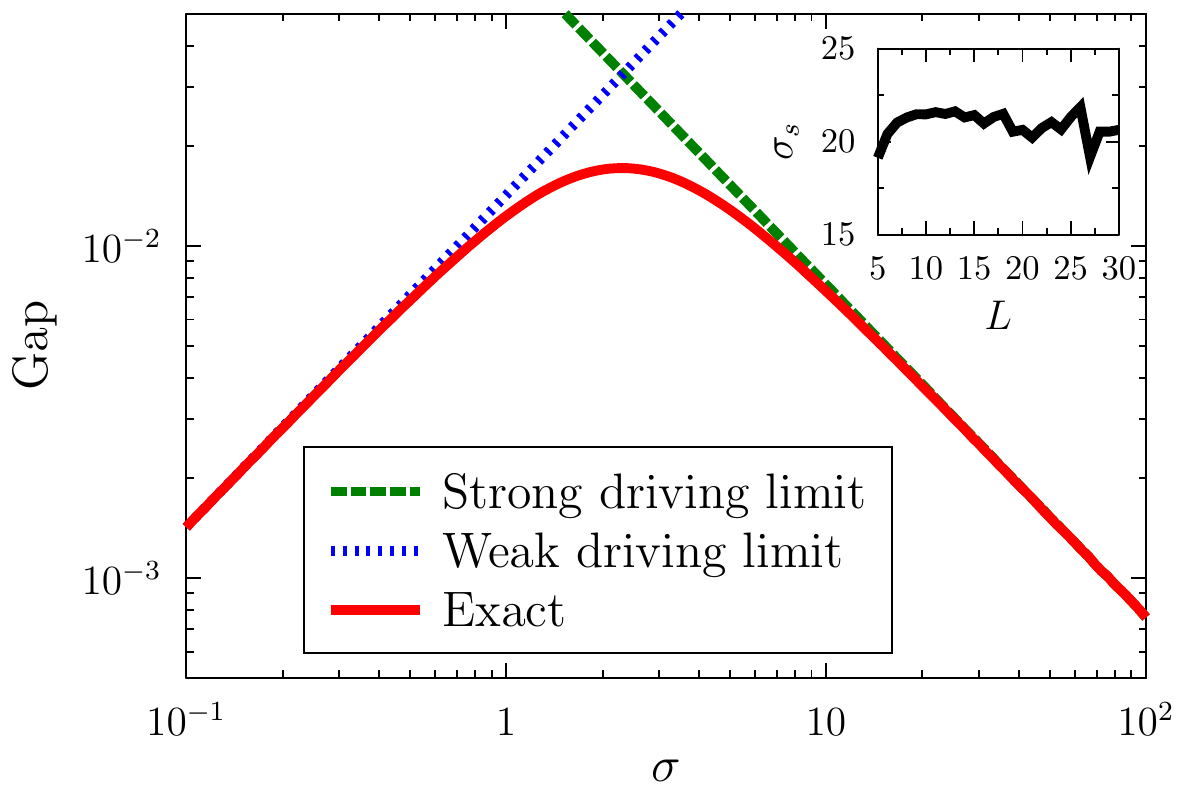}
  \caption{Liouvillean gap for a controllable chain of $L=10$ as a function 
  of the noise strength $\sigma$. Exact numerical results are obtained with $q=1$. 
  Strong driving limit corresponds to Eq.~\eqref{e:gapfinal}, while the weak 
  driving limit is from Eq.~\eqref{e:gapA}. Inset: noise strength 
  $\sigma_s$ as a function of $L$ such that, for $\sigma>\sigma_s$, 
  the relative error between the exact gap and the strong coupling 
  estimate is smaller than $1\%$. 
  }
  \label{fig:eff}
\end{figure}
In Fig.~\ref{fig:eff} we study the Liouvillean gap and show that the theoretical predictions 
of the strong and weak driving limits are very accurate in their respective limit of validity.
Moreover, we found that the accuracy of the strong driving limit is not affected by the length 
of the chain. This is shown indeed in the inset Fig.~\ref{fig:eff} where one observes 
an almost constant behaviour as a function of $L$. 
\begin{figure}[t]
  \centering
  \includegraphics[width=0.40\textwidth]{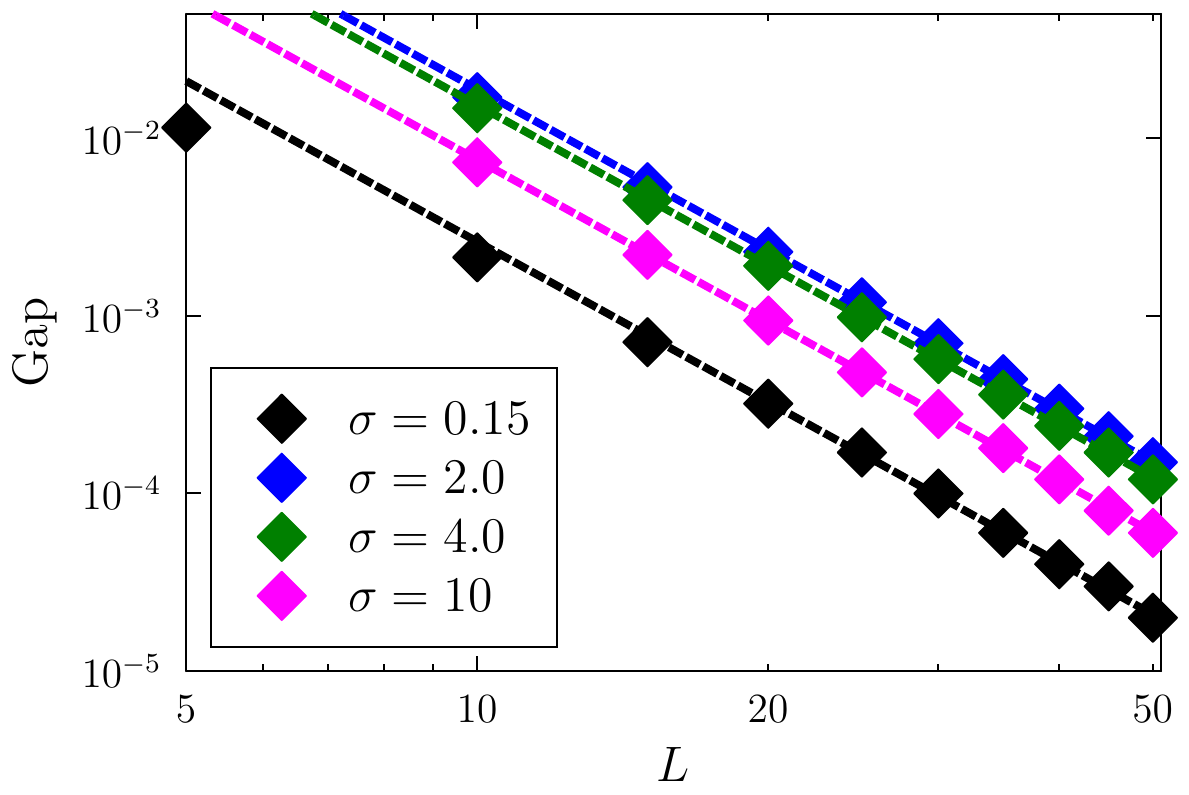}
  \caption{Scaling of the Liouvillean gap obtained numerically for $q=1$ as a function 
    of $L$ and for different values of $\sigma$. Solid lines corresponds to fitting 
    functions $\propto L^{-3}$. 
  }
  \label{fig:scaling}
\end{figure}
In Fig.~\ref{fig:scaling}, on the other hand, we show that the Liouvillean gap scales as 
$L^{-3}$ for different values of $\sigma$. This scaling 
has been predicted in the strong and weak 
driving limits by Eqs.~\eqref{e:gapfinal} and \eqref{e:gapA}. However, 
Fig.~\ref{fig:scaling} shows that such scaling is valid also for $\sigma\approx2$ where 
neither the strong nor the weak coupling limit holds (compare e.g. the values 
of Fig.~\ref{fig:scaling} and Fig.~\ref{fig:eff}).
\begin{figure}[t]
  \centering
  \includegraphics[width=0.35\textwidth]{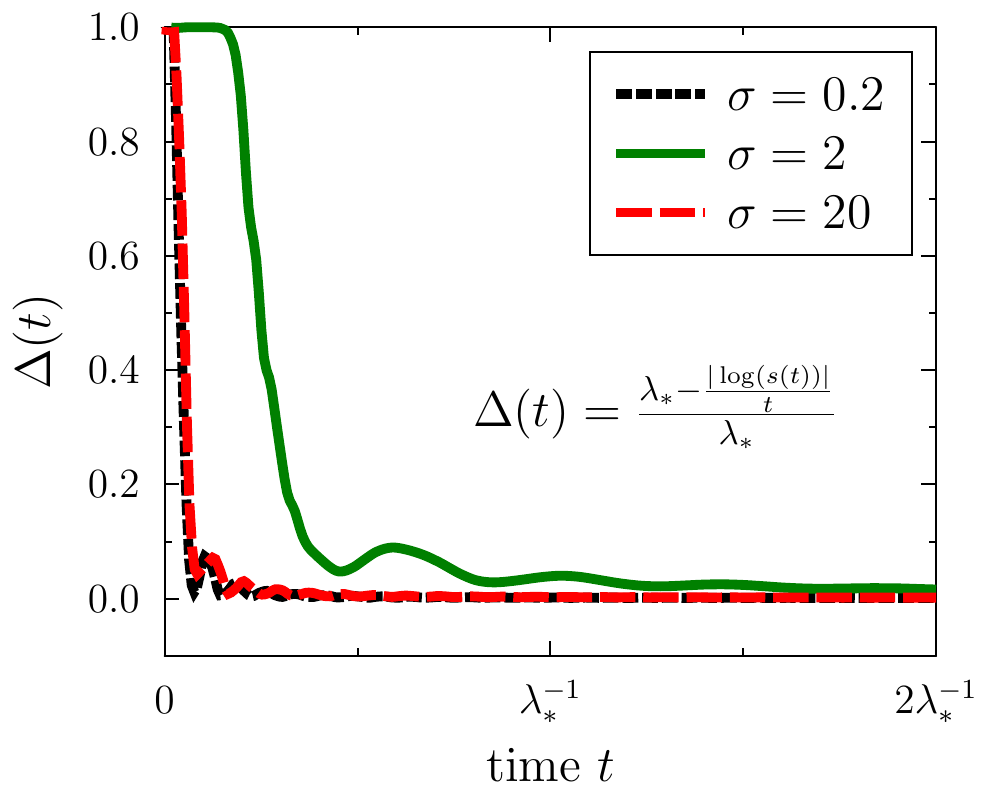}
  \caption{ Convergence of the singular value $s_*(t)$ of 
    $e^{t \mathcal L} e^{t \mathcal L^\dagger}$ to $e^{-\lambda_* t}$ 
    where $\lambda_*$ is the Liouvillean gap. The relative error $\Delta(t)$ 
    between $\lambda_*$ and $-t^{-1}\log s(t)$ is 
    plotted for the different values of 
    $\sigma$, the time axis is rescaled between 0 and $2\lambda_*^{-1}$. In the simulations
    $L=10, q=1$. 
  }
  \label{fig:singv}
\end{figure}
In Fig.~\ref{fig:singv} we study the relationship between the Liouvillean gap and the 
gap $s^*(t)$ in the singular values of $e^{t \mathcal L_q}$ which is 
a good estimate of the convergence time (see section \ref{s:convtime}). 
As expected, both in the strong and weak coupling limit the $s^*(t)$ converges 
to $e^{-\lambda^* t}$ much earlier than mixing time-scales. Therefore, in these
regimes, one finds that the convergence time is basically $1/\lambda^*$. 
On the other hand, for $\sigma=2$ the matching between 
$e^{-\lambda^* t}$ and $s^*(t)$ only happens at longer times. Therefore, as 
expected from the analysis of Section~\ref{s:convtime}, in this 
regime there is a correction to the mixing time due to the norm of the 
left and right eigenvectors. 
Nonetheless, similarly to the Liovillean gap, our numerical simulations for 
small values of $L$ and $q$ show that also the singular value gap is independent 
on $q$ over the whole range of $\sigma$. Therefore, we argue that it may be a general 
feature of this model that the resulting convergence time is independent on $q$. 

\begin{figure}[t]
  \centering
  \includegraphics[width=0.22\textwidth]{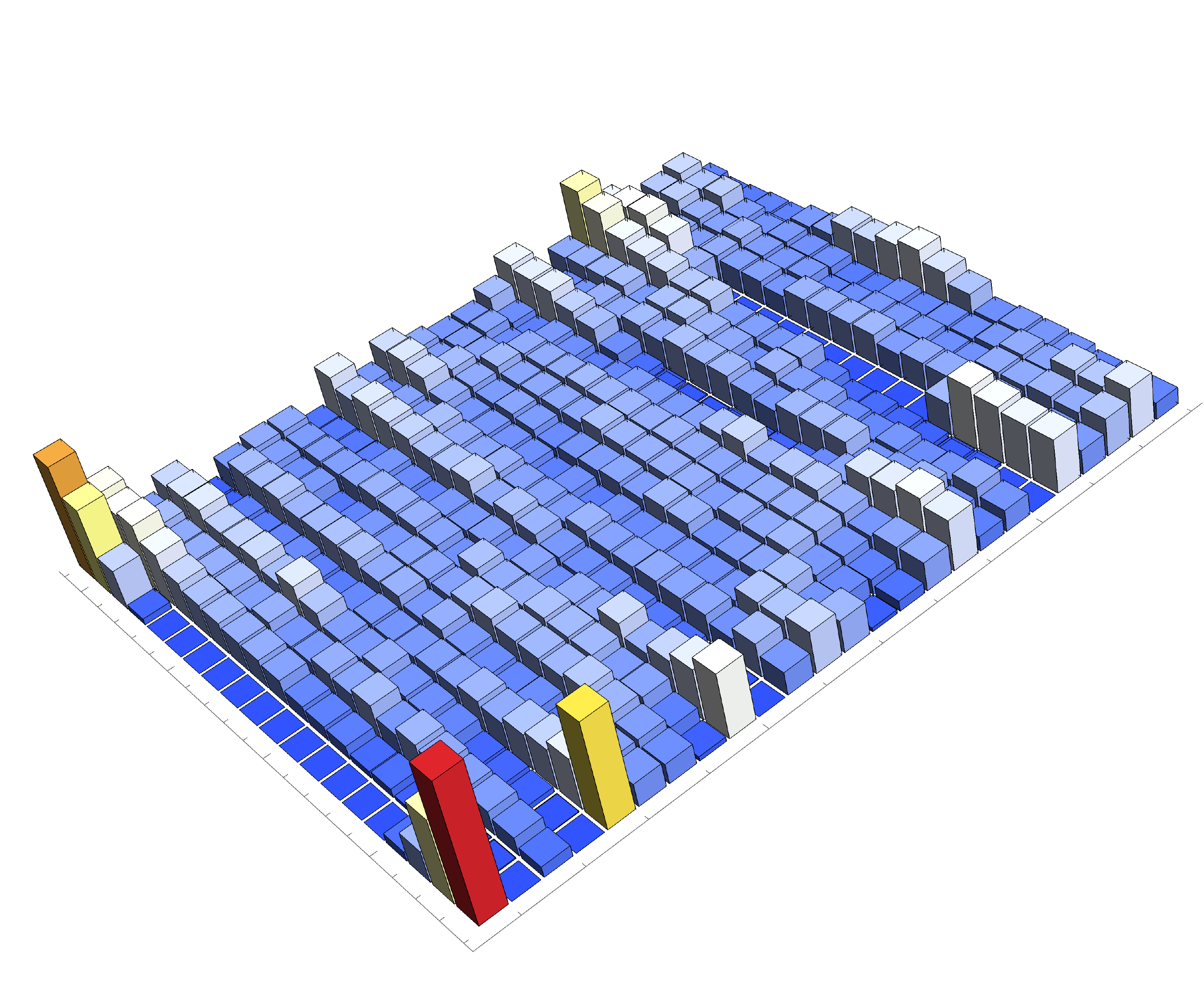}
  \includegraphics[width=0.22\textwidth]{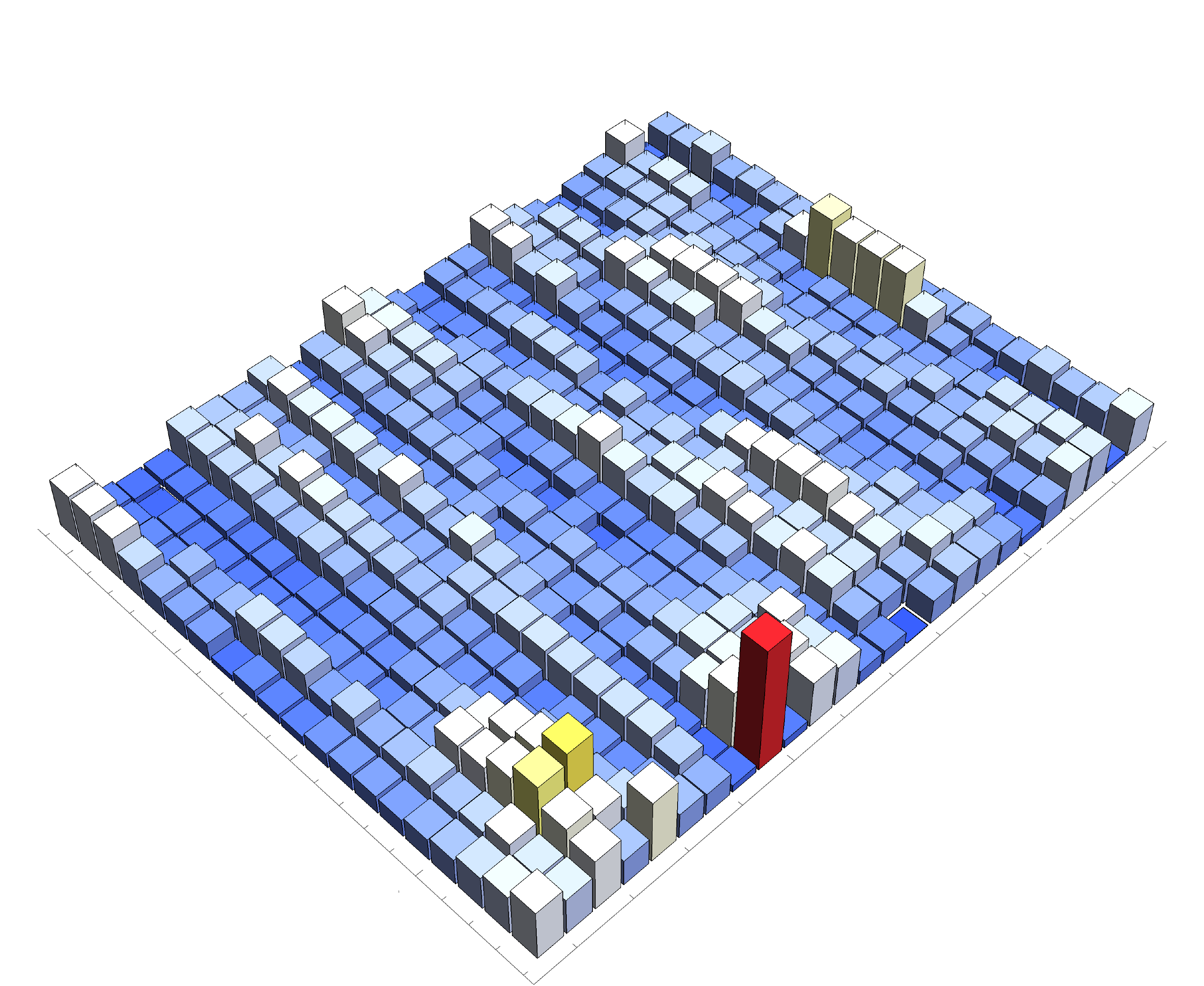}
  \\
  \includegraphics[width=0.36\textwidth]{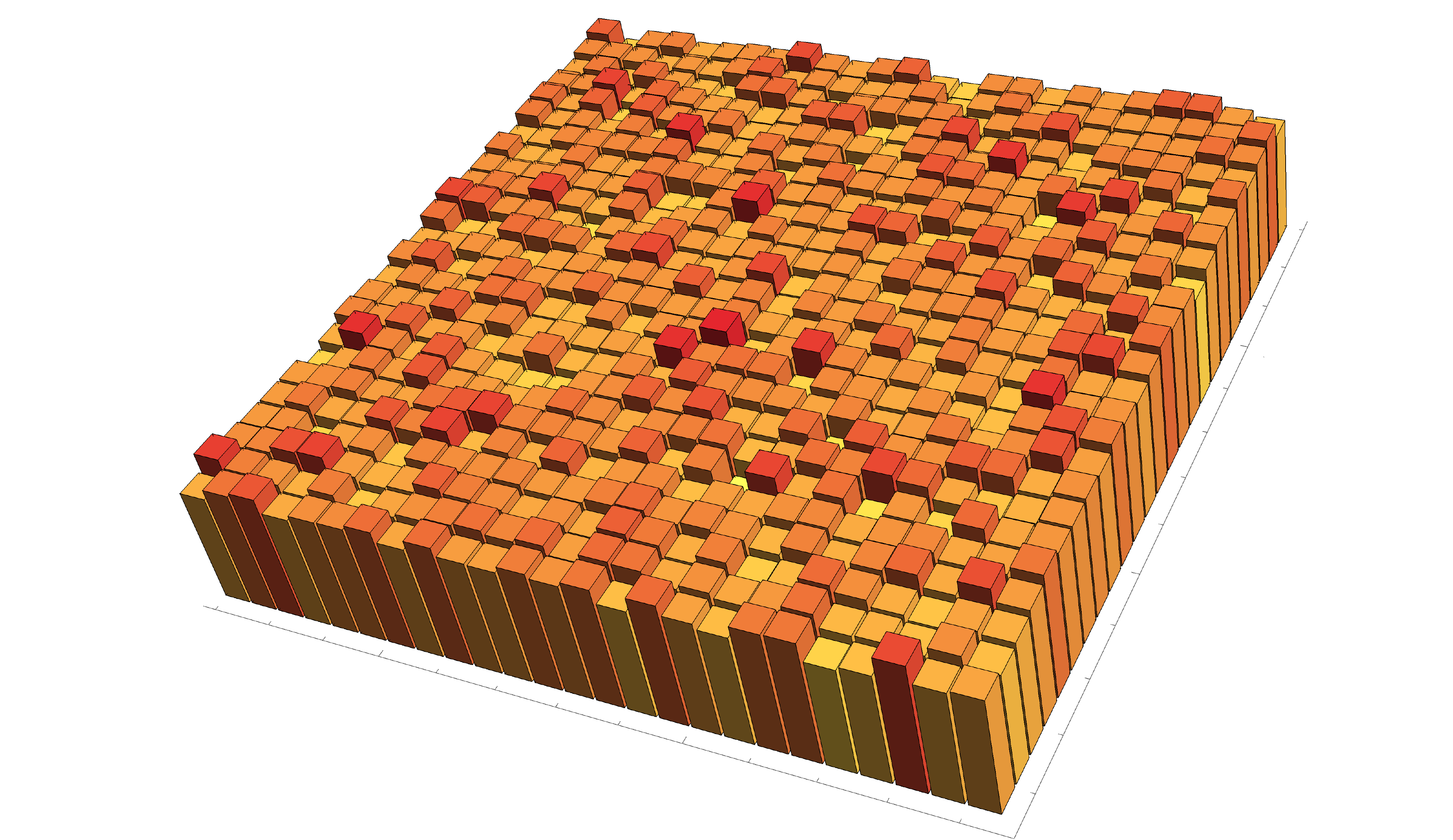}
  \caption{
    Uniformity check of generated random unitaries. We considered the time evolution
    of a driven $L=5$ chain with random fields \eqref{e:fourier}, 
    where $K=100$, $g_k$ is sampled uniformily in [-0.5,0.5], while $\phi_k$ and $\omega_k$ are 
    sampled in $[-L,L]$. The statistics is done with $10^4$ independent realizations. 
    The discrete histrogram is computed according to the decomposition 
    \cite{spengler2012composite} as described in the main text. 
    (a) Non-controllable case where noise is applied on the central site for a time 
    $t=25$. 
    (b) Controllable case where the noise is applied on the first site for a short 
    time $t=5$. 
    (c) Controllable case where the noise is applied on the first site for a long
    time $t=55$. 
  }
  \label{fig:bricks}
\end{figure}

Finally we consider a stochastic simulation of the evolution of a controllable chain with
random fields: we generate several random driving functions \eqref{e:fourier} and, 
for each function, we calculate the corresponding unitary evolution and then study the 
statistics of the generated unitary matrices. To test whether the resulting distribution 
approximates the Haar measure we decompose each unitary into the $L^2$ angles 
introduced in Ref.~\cite{spengler2012composite}. Using a simple reparametrization of 
these angles one can write the Haar measure as 
\begin{align}
dU(\varphi_1,\dots,\varphi_{L^2})= \prod_{j=1}^{L^2} d\varphi_j ~,
\end{align}
namely as a uniform distriution of the angles $\varphi_j$ in the 
range $[0,2\pi]$. Therefore, testing whether the resulting distribution approximates a Haar
measure is equivalent to testing whether the angles $\varphi_j$ are distributed as a multinomial
uniform distribution. In Fig.~\ref{fig:bricks} we do a simple test to verify the distribution
of the angles $\varphi_j$: we divide the interval $[0,2\pi]$ into 25 bins and plot,
as a 3D histrogram, the matrix whose elements $(i,j)$ are the number of 
times that the angle $\varphi_i$ is found in the $j$-th bin.
As Fig.~\ref{fig:bricks} shows, the distribution of the unitary matrices 
is far from uniform both in the noncontrollable case and 
in the controllable case after a short time (upper panel). 
Nonetheless, in spite of the finite number of 
samples, after a long time ($t\approx 55$) in the controllable case the angles'
distribution is almost flat (lower panel), thus showing that the resulting unitary matrices are 
approximately distributed according to the Haar measure.

\section{Other applications }
\label{s:appl}
\subsection{Multi-point correlation functions}
\label{s:applmp}
Here we discuss some direct applications, beyond $q$-design, of the main findings of our paper. 
In boson sampling experiments the output probability is proportional to 
$|{\rm per}(\tilde{U})|^2$, being ${\rm per}(\tilde U)$ the matrix permanent of the 
$q\times q$ matrix $\tilde U$, where $\tilde U$ is 
built from some columns and rows of a $L\times L$ Haar-uniform matrix $U$
\cite{broome2013photonic,spring2013boson,crespi2013integrated}. Therefore
\begin{align}
  |{\rm per}(\tilde U)|^2 &= \sum_{\sigma,\sigma'} \prod_{i,j=1}^q 
  \tilde{U}_{i,\sigma(i)} 
  \tilde{U}^*_{j,\sigma'(j)} 
  \\ &= \Tr\left[ U^{\otimes q,q} \mathcal K_{\rm b.s.}\right]~,
  \label{e:per}
\end{align}
where $\sigma,\sigma'$ are permutations in the symmetric group $S_q$, 
$\mathcal K_{\rm b.s.}$ is a suitable  index contraction operator and $U^{\otimes q,q} =
U^{\otimes q}\otimes (U^{\otimes q})^*$ as in Eq.~\eqref{e:expander}. 

A similar expression arises in the evaluation of multi-point correlation 
functions in quasi-free particle-preserving bosonic and fermionic models. 
If $U$ is the $L\times L$ one particle evolution
matrix from time $0$ to time $t$ and $a_j(t)=\sum_k U_{jk} a_k(0)$, then because 
of the Wick's theorem
\begin{align}
  \langle 
  a^\dagger_{i_1}(t)a_{j_1}(t)\dots
  a^\dagger_{i_q}(t)a_{j_q}(t)
  \rangle 
  = \Tr\left[ U^{\otimes q,q} \mathcal K_{\rm m.p.}\right]~,
  \label{e:multipoint}
\end{align}
where $\mathcal K_{\rm m.p}$ depends on the initial two-point correlation functions
$\langle a_i^\dagger(0) a_j(0)\rangle$. Expressions like \eqref{e:multipoint} arise
also in XY spin chains, which can be mapped to a quasi-free fermionic model via the
Jordan-Wigner transformation \cite{lieb1961two}. For instance, the driven XY model 
\begin{equation}
H_{\rm XY}(t)=\frac12\left[\sum_j(\sigma^x_j\sigma_{j+1}^x +
\sigma^y_j\sigma_{j+1}^y) + g(t) \sigma_1^z\right]~,
\end{equation}
can be mapped, 
in the single-particle subspace, to the driven quantum walk of Section \ref{s:chain}. 
Calling $U$ the resulting single-particle evolution,
then in {\it any subspace} long-range spin 
operators $S^\alpha_i S^\beta_j$, for $\alpha,\beta\in\{x,y\}$ can be written as 
a combination of fermion strings as in  
\eqref{e:multipoint} where $q=|i-j|$ for $i\neq j$. Therefore, 
with a suitable $\mathcal K_{\rm XY}$ that depends on the initial correlations, 
one can write the dynamical long-range correlations between spin operators in an XY chain as
\begin{align}
  \langle 
  S^\alpha_i(t) S^\beta_{i+q}(t)
  \rangle 
  = \Tr\left[ U^{\otimes q,q} \mathcal K_{\rm XY}\right]~,
  \label{e:xy}
\end{align}
for $\alpha,\beta\in\{x,y\}$. 
Similarly, $\langle S^z_i(t) S^z_{j}(t) \rangle 
  = \Tr\left[ U^{\otimes 2,2} \mathcal K^{zz}_{\rm XY}\right]$.

In all the above cases we can bound the convergence of the random dynamics 
to the values expected from the Haar distribution. Indeed, for any $\mathcal K$
\begin{align}
  \left|\Tr\left[\left(\mathbb{E}_UU^{\otimes q,q} -
    \int  U^{\otimes q,q} \,\mu_{\rm Haar}(dU)\right)\mathcal K\right]\right| < 
    e(\mu_U,q) \|\mathcal K\|_1~,
\end{align}
where we used \eqref{e:expander}. Thanks to the analysis of Section \ref{s:convtime},
and since the gap \eqref{e:gapfinal} for the controllable quantum walk is independent
on $q$, one can then bound the expected errors in all the above cases. 
For boson sampling experiments, this shows how the error depends on the number $q$ of bosons, while
for XY spin chains, it shows how the error decays as a function of the distance $q$ between
spins.

\subsection{Estimation of the control time}
\label{s:applcontr}
We show here that the mixing time, which is easy to compute especially for $q=1$, 
can give an estimation of the control time. Fixing $H$ and $V$, for how long does one
have to drive the system in order to achieve a generic target gate? 
If after the time $T^*_{ex}$ the random evolutions are Haar-randomly distributed, then the control
time to obtain a certain gate $U$ satisfies $T_c(U)<T^*_{\rm ex}$. However, for approximate
$q$-design, $T^*$ provides only a rate of convergence, rather than a sharp bound. 
This results into an error, which may also be due to the fact that the target gate $U$ 
is not achievable yet at time $T^*$. However, after a time $\tau T^*$ this error probability 
exponentially decreases as a function of $\tau$. 
We can thus regard $T^*$ as an estimation for $T_c$. An estimation of the 
mixing time $T^*$ can be 
easily obtained for any choice of $H$ and $V$ via the inverse of the gap $\lambda^*$, 
which depends on $\sigma$ 
(see e.g. Fig.~\ref{fig:eff}). Since $T_c$ does not involve any specific properties 
(amplitudes, frequencies) of the pulse, one has to compare it with 
$T^*_{\rm min} = \min_\sigma T^*(\sigma)\simeq T^*(\sigma\simeq2.5)\approx 0.055 \,L^3$. 

In order to estimate $T_c$ we perform a numerical experiment with the QuTip quantum
control package \cite{johansson2012qutip}. We consider the model \eqref{e:H1} and, for 
each length $L=10,\dots,20$, 
we generate a Haar-random unitary $U$ and find the time $T_c$ as the minimal
time for which the program converges.
We find that $T_c$ obtained in this way scales as $T_c\approx 0.069 \,L^3$.
This shows two remarkable facts: (i) the values of $T_c$ and $T^*_{\rm min}$ are very close
for $L=10,\dots,20$; (ii)
both $T_c$ and $T^*_{\rm min}$ exhibit the same 
scaling with the length $L$, so it is expected that this close relationship is maintained 
also for larger $L$. 
In view of our findings, one can find an empirical upper bound on $T_c$ as 
$3 T^*_{\rm min}/2$.

\section{Conclusions and perspectives}
\label{s:concl}

In this paper we have studied the quantum dynamics resulting from a stochastic driving
of quantum many-body systems, and we have answered the following questions: when, and how
rapidly, the dynamics of a driven quantum system is equivalent to 
to a fully uniform random evolution, namely 
under unitaries sampled from the Haar measure. 
The first major finding is that, when the system is {\it fully controllable} and 
the stochastic signal has finite correlation time, then its random dynamics 
converges to the Haar distribution in the ``long time'' limit. 
The second major result is about the estimation of the driving time $T^*$: this is done 
by studying the deviations from the Haar distribution using the framework of 
approximate $q$-design, and using second-quantization to map the problem 
into the estimation of the mixing time in an open quantum many-body Liouvillean with
$2q$ virtual particles. 

We have performed a thorough analysis of 
the Markovian limit (e.g. white noise) using tools from 
the theory of dynamical semigroups,
and we found upper bounds on $T^*$ in terms of the gap of the Liouvillean operator.
We studied the mean field solution of the resulting many-body 
model, which predicts a constant Liouvillean gap as a function of $q$,
and we have shown its limitations via symmetry breaking arguments. 
Nonetheless, we found that the mean-field predictions are correct in a 
wide variety of different numerical studies, obtained with random choices 
of $H$ and $V$, and match with the analytic 
solution of a particular model, namely a
one dimensional system with strong control on one of its boundaries. 
The latter analytic solution 
has been obtained by mapping the effective Liovillean to an 
exactly solvable model, and then using Bethe-Ansatz techniques to 
explicitly 
show that the excited states with smallest gap are built from 
unpaired quasi-particles, as in the mean field treatment. 
We have then corroborated our predictions with numerical simulations, 
putting strong evidence that the considered one-dimensional model 
provides a quantum expander with a constant mixing time as a function
of $q$.
Therefore, our results show that certain driven physical systems can 
provide a significant advantages over 
random quantum circuits where the mixing time increases polynomially 
as a function of $q$ \cite{brandao2012local}.

The results presented in this paper have many applications. The first one,
already discussed, is a physically motivated approach to generate 
pseudo-uniform random unitary operations, which have many applications 
in quantum information processing protocols. 
The one-dimensional system 
that is extensively analyzed in this paper is motivated by the 
recent experiments with integrated photonic circuits
\cite{perez2013coherent,pitsios2016photonic}, where random unitary 
operations have been used in the first small-scale 
experimental observations of boson sampling 
\cite{broome2013photonic,spring2013boson,crespi2013integrated}. 
The results presented in this paper enable the implementation
of random operations in integrated photonic chips that, 
being based on noisy quantum walks rather than carefully designed 
multi-mode beam splitters and phase shifters, are much simpler 
to fabricate for a larger number of modes. 
Therefore, our results provide a new avenue to prove quantum supremacy
in boson sampling experiments. 

Moreover, we have considered other applications, such as the dynamics of 
correlation functions in an XY spin chain, and the estimation of the control 
time $T_c$, one of the major open problems for quantum control. 
Given a target unitary $U$ and the physical interactions described by $H_0$ and $V$, 
how can we chose $T_c$ such that $U$ is
achievable by driving the system for a time $T_c$? 
With numerical experiments, performed on $L$-site chains, 
we found that both $T_c$ and $T_*$ are very close for $L=10,\dots,20$, 
and both scale as $L^3$. Hence, 
the mixing time $T_*$ under random signals provides an easily  computable estimation of 
$T_c$, for any $H_0$ and $V$. 

Finally, there are several applications in quantum many-body physics, where
the interplay between quantum many-body effects and noise is 
currently a subject of intensive study in many area, such as 
spin glass \cite{mezard1987spin}, the fast scrambling of quantum information
\cite{hayden2007black,sekino2008fast}, and many-body localization 
\cite{pal2010many,ponte2015many}. 
The explicit one dimensional model discussed in Section \ref{s:chain} is 
a single-particle model, where many-body physics arises due to unitary 
$q$ design, which introduces $2q$ virtual particles.  An interesting future
perspective is the study of random driving in physical interacting many-body 
systems (e.g. interacting spin systems and/or cold atoms optical lattices).
In fact, the competition between physical many-body effects, and those arising 
from the unitary design, may give rise to novel states of matters and 
phase transitions \cite{medvedyeva2016exact,vznidarivc2015relaxation,banchi2014quantum,diehl2011topology,prosen2008quantum}, produce large amount of entanglement 
\cite{nahum2016quantum}, and give new insights into the process of thermalization and 
equilibration \cite{eisert2015quantum}. 
Haar-random quantum states are known to have, typically, an extensive amount of entanglement 
\cite{hayden2006aspects}. 
Since we have shown that any controllable quantum system
converges to a maximally mixing dynamics, 
the real time dynamics will be very hard to simulate numerically in the many-body settings,
because of the large amount of entanglement involved. 
Nonetheless, the controllability requirement provides a sufficient 
algebraic method to infer, {\it a priori}, whether a randomly 
driven condensed matter system 
is expected to produce lots of entanglement in the long time limit.

\begin{acknowledgements}
  The authors thank 
  S. Bose, 
  E. Compagno, 
  F. Falceto, 
  J. Links, 
  A. Marcus, 
  L. Maccone, 
  S. Maniscalco,
  E. Mukhin, 
  S. Paesani, 
  R. Santagati, 
  S. Severini, 
  A. Werner,
  R. Zeier, 
  and 
  Z. Zimbor\'as
for interesting discussions. 
L.B. has received funding for this research from the European Research Council under the European Union’s Seventh Framework Programme (FP7/2007-2013)/ERC Grant agreement No. 308253 PACOMANEDIA. MJK was supported by the Villum foundation. 
D.B. acknowledges support from the EPSRC grant No. EP/M01634X/1. 
\end{acknowledgements}



%

\appendix

\section{Gaussian harmonic pulses}
\label{s:markov}

To simplify the theoretical description, in this section we consider only 
$q=1$ and 
call $\mathcal E_t$ the quantum
channel resulting from the average evolution of the quantum system 
\begin{align}
  \mathcal E_t[\rho] = \mathbb{E}\left[
\mathcal T \re^{{-}\ri \int_0^{t}\mathcal H(s)\, ds}\,\rho\,
\mathcal T \re^{\ri \int_0^{t}\mathcal H(s) \,ds}\right]~. 
\label{e:Et1}
\end{align}
Extensions to higher values of $q$ is straightforward. As described in 
section \ref{s:design}, we now make two assumptions, namely that 
$g(t)$ is Gaussian and harmonic, where
$\mathbb{E}[g(t+s)g(t)] = c(s)$ is independent on $t$ and 
$\mathbb{E}[g(t)]=0$. In view of these assumptions, 
we can simplify \eqref{e:Et1} by expanding the exponentials into the Dyson series, 
then using the Wick's theorem to decompose the expectation values and finally 
resumming the series. The result in the interaction picture is then 
\cite{ishizaki2009unified,banchi2013analytical} 
\begin{align}
  \mathcal E^{(I)}_t[\rho^{(I)}] &= \mathcal T \re^{{-}\int_0^t ds \mathcal W_s} \rho^{(I)}~,
\label{e:Et2}
  \\ \mathcal W_s \rho &= \int_0^s  c(s-s')\, [V^{(I)}(s), \,[V^{(I)}(s'), \,\rho]]\, ds' ~, 
  \nonumber
\end{align}
where $(I)$ refers to the interaction picture with respect to $H$. If the correlation time
is finite then there exist a suitably large $T$ such that 
$T c(T s) \simeq \frac\sigma2 \delta(s)$ where $\delta$ is the Dirac delta function ans 
$\sigma$ is a constant. In the long-time limit one finds that 
\begin{align*}
  \mathcal E^{(I)}_{t}[\rho^{(I)}] \simeq  \mathcal T 
  \exp\left({-}c \int_0^t [V^{(I)}(s), \,[V^{(I)}(s), \,\cdot]]\, ds\right) \rho^{(I)} ~, 
\end{align*}
when $t>T$, namely in the Schr\"odinger picture
\begin{align}
  \mathcal E_{t}[\rho] &\simeq  \re^{-t \mathcal L} \, \rho
  \nonumber
  \\ 
  \mathcal L \rho &= -\ri[H, \rho] - \frac\sigma2[V, [V, \rho]]~. 
  \label{e:lindblad1}
\end{align}

\section{Semigroup convergence times}
\label{s:smeiconv}

There exist several measures to estimate convergence of a semigroup of completely positive trace preserving (CPTP)  maps. The one with the most natural operational interpretation is trace norm convergence, as it reflects the likelihood that the time evolved state can be distinguished from the stationary state at a given time $t$. 
 \begin{equation}
 \sup_{\rho}||e^{t\mathcal{L}}(\rho)-T_\infty(\rho)||_1\leq \epsilon(t),
 \end{equation}
where $T_\infty=\lim_{t\rightarrow\infty} e^{t\mathcal{L}}$, and $\epsilon(t)$ is the distinguishability error. 
A less stringent convergence requirement is to ask whether $e^{t\mathcal{L}^q}$ is an expander for a given value of $t$. Then, we want to estimate

\begin{equation}
||e^{t\mathcal{L}}-T_\infty||_{2\rightarrow 2}=||e^{t\hat{\mathcal{L}}^q}-\hat{T}_\infty||_\infty,\label{eqn:infnorm}
\end{equation}
where a hat indicates that the CPTP maps are represented as channels (see Ref. \cite{Wolfnotes} for more details on the representation of channels). Trace norm convergence and 'spectral convergence' are related, by noting that 
\begin{equation}
||e^{t\mathcal{L}}-T_\infty||_{2\rightarrow 2}\leq||e^{t\mathcal{L}}-T_\infty||_{1\rightarrow 1}\leq d^{2}||e^{t\mathcal{L}}-T_\infty||_{2\rightarrow 2},\label{normbounds}
\end{equation}
where $d$ is the dimension of the Hilbert space, and recalling that $||e^{t\mathcal{L}}-T_\infty||_{1\rightarrow 1}=\sup_{\rho}||e^{t\mathcal{L}}(\rho)-T_\infty(\rho)||_1$.

In order to estimate the above norms it is important to recall the spectral properties of quantum dynamical semigroups. The spectrum of a Liouvillian $\mathcal L$ has non-positive real part, and there always exists at least one eigenvalue of magnitude zero, corresponding to a stationary state of the semigroup: $\mathcal{L}(\rho)=0$. The rest of the spectrum comes in complex conjugate pairs. The Liouvillian is called unital if it annihilates the identity $\mathcal{L}(\openone)=0$. The Liouvillian in Eq. (\ref{e:lindblad}) has this property. A unital Liouvillian is called reversible if $\hat{\mathcal{L}}=\hat{\mathcal{L}}^\dag$, in which case its spectrum is real. Unfortunately, Eq. (\ref{e:lindblad}) is not reversible. Convergence of a non-reversible semigroup is governed by the singular values of $e^{t\mathcal{L}}$ rather than its eigenvalues. The singular spectrum of   $e^{t\mathcal{L}}$ is equal to the spectrum of $\sqrt{ e^{t\hat{\mathcal{L}}}e^{t\hat{\mathcal{L}}^\dag}}$. 

It is not difficult to see that the $2\rightarrow 2$ norm is related to the singular spectrum. Let $s_j(t)$ be the singular values of $e^{t\mathcal{L}}$, ordered from largest to smallest. The largest has magnitude one. 
We know that asymptotically $s^*(t)=e^{t {\rm Re}[\lambda^*]}$, where now $\lambda_j$ are the eigenvalues of $\mathcal{L}$ written in decreasing (real part) order, and $\lambda^*$ is the \textit{gap} of $\mathcal{L}$; i.e. the smallest (in magnitude) non-zero real part of any eigenvalue of $\mathcal{L}$. To see this, note that, assuming it has no Jordan blocks, the Liouvillian can be written in its spectral decomposition as 
\begin{equation}
\mathcal{L}(\rho)=\sum_j \lambda_j L^\dag_j {\rm tr}[R_j \rho],
\end{equation}
where $R_j,L_j$ are a bi-orthonormal basis of operators: i.e. ${\rm tr}[L^\dag_j R_k]=\delta_{jk}$. Importantly, the norm of any given $L_j,R_j$ can be large, which prevents us from getting any rigorous (universal) bounds between the singular values and the eigenvalues. Then, 
\begin{eqnarray}
  \nonumber
||e^{t\mathcal{L}}-T_\infty||_{2\rightarrow 2} &=& \sup_{\psi} (\sum_{j:\lambda_j\neq0} e^{2t{\rm Re}[\lambda_j]}|\langle R_j|R_j\rangle|~|\langle\psi| L_j\rangle\langle L_j|\psi\rangle|)^{1/2}\\
&\approx_{t\rightarrow\infty}& e^{t\lambda^*}|\langle R_j|R_j\rangle|~|\langle\psi| L_j\rangle\langle L_j|\psi\rangle|)^{1/2}.\label{eqn:appA}
\end{eqnarray}
Hence, for very large $t$, the convergence is governed by the gap, and  $s^*(t)\rightarrow e^{t\lambda^*}$. In principle we do not know at what scale $e^{-t\lambda^*}\gg |\langle R_j|R_j\rangle|~|\langle\psi| L_j\rangle\langle L_j|\psi\rangle|$.

We argue in the main text, that for the specific model of a controllable quantum walk, the prefactors do not contribute to the asymptotics in the weak or strong coupling limits.

\section{Weak driving limit}
\label{s:weak}

A convenient approximation for the long-time dynamics in the weak coupling limit 
$\sigma\ll 1$ is the rotating wave
approximation (RWA) \cite{fleming2010rotating}. 
We consider the case $q=1$ and assume that $V$ is a matrix of real numbers and 
call $\mathcal D=-\sigma \mathring{V}^2/2$ the dissipative part in \eqref{e:Lvec}. 
Going to the interaction picture with respect to the
Hamiltonian part one finds that $\dot\rho_I(t)=\mathcal D_I(t)\rho_I(t)$ where 
in the eigenbasis of $H=\sum_j \omega_j \ket{\omega_j}\bra{\omega_j}$ it is 
\begin{align}
\bra{\omega_i\omega_j}\mathcal D_{I}(t)\ket{\omega_k\omega_l}=& -\frac{\sigma}2
e^{-it(\omega_{ij}-\omega_{kl}} \mathcal R_{ijkl},
\\
R_{ijkl}= 
&
\bra{\omega_i}V^2\ket{\omega_k}\delta_{jl}+
\bra{\omega_j}V^2\ket{\omega_l}\delta_{ik}
\nonumber \\&
-2
\bra{\omega_i}V\ket{\omega_k} \bra{\omega_j}V\ket{\omega_l}~,
\end{align}
where $\omega_{ij}=\omega_i-\omega_j$. 
The rotating wave approximation consists in neglecting all the terms where
$\omega_{ij}\neq\omega_{kl}$, because for  large $t$ they are highly
oscillating and average out:
\begin{align}
    \bra{\omega_i\omega_j}\mathcal D^{\rm RWA}\ket{\omega_k\omega_l}=
    R_{ijkl} \delta_{\omega_{ij},\omega_{kl}}.
    \label{e:Drw}
\end{align}
This approximation is expected to hold when 
\begin{align}
  t\gg \max_{\omega_{ij} \neq \omega_{kl}} (\omega_{ij}-\omega_{kl})^{-1}~.
  \label{e:RWAlarget}
\end{align}
RWA is related to degenerate perturbation theory. Indeed, the 
unperturbed ($\sigma=0$) eigenvalues of \eqref{e:Lvec} are given by 
$\ket{\Phi^{(0)}_{ij}} = \sum_{kl}\delta_{\omega_{ij},\omega_{kl}} \alpha_{kl}^{ij}
\ket{\omega_k\omega_l}$ with eigenvalue 
$-i\omega_{ij}$. From degenerate first order perturbation theory we know that, 
for small $\sigma$, the eigenvalues of Eq. \eqref{e:Lvec} are 
obtained by diagonalising $\mathcal D^{\rm RWA}$,  which is block diagonal where
each block acts on different degenerate subspaces. The eigenvectors of 
$\mathcal D^{\rm RWA}$ 
provide the matrices $\alpha^{ij}_{kl}$. Note that since 
$\mathcal D^{\rm RWA}$ is Hermitian, the states $\ket{\Phi_{ij}^{(0)}}$ 
form an orthonormal basis which depends both on $H$ (from the basis $\ket{\omega_k}$)
and $V$ (via the diagonalization of $\mathcal D^{\rm RWA}$). 
Moreover, the real eigenvalues 
$\Delta_{ij}$ of \eqref{e:Drw} provide the first order 
correction to the eigevectors of Eq. \eqref{e:Lvec} that, to the first 
order in $\sigma^{-1}$ are $-i\omega_{ij} + \Delta_{ij}$. The Liovillean 
gap is given by the minimum non-zero value of $-\Delta_{ij}$. Similarly one finds the
correction to the (right) eigenvector 
\begin{align}
  \ket{\Phi_{ij}^{(1)}} &=
  \ket{\Phi_{ij}^{(0)}} -\ri 
  \sum_{\substack{kl \\ \omega_{kl}\neq\omega_{ij}}}
  \ket{\Phi_{kl}^{(0)}}
  \frac{\bra{\Phi_{kl}^{(0)}}\mathcal D 
  \ket{\Phi_{ij}^{(0)}} }{\omega_{kl}-\omega_{ij}}
\nonumber \\& 
\simeq
\re^{\mathcal S^{\rm RWA} } \ket{\Phi_{ij}^{(0)}} 
  ~,
  \label{e:RWAev}
\end{align}
where 
\begin{align}
\mathcal S^{\rm RWA}=-\ri
 \sum_{\substack{klmn \\ \omega_{kl}\neq\omega_{ij}}}
  \ket{\Phi_{kl}^{(0)}}
  \frac{\bra{\Phi_{kl}^{(0)}}\mathcal D 
  \ket{\Phi_{mn}^{(0)}} }{\omega_{kl}-\omega_{mn}}
  \bra{\Phi_{mn}^{(0)}}~.
  \label{e:Srwa}
\end{align}
Since 
$\mathcal S^{\rm RWA}$ is a Hermitian operator the new vectors 
in \eqref{e:RWAev} do not form an orthonormal basis. 

We now focus on the the the chain discussed in Section \ref{s:chain} 
where $\omega_{k_j}=2\cos k_j$, $k_j=\pi
j/(L+1)$, $V^2=V$ and we call
$W_{ij}=\bra{\omega_i}V\ket{\omega_j}=\frac{2}{L+1}\sin k_i\sin k_j$. 
To simplify the equations we use the compact notation $\ket i\equiv
\ket{\omega_i}$ and we use $c=1$, namely we assume that the controlled site is the first
one. 
We note that the resonance condition 
$\omega_i-\omega_j= \omega_k-\omega_l$ is achieved in three different cases: 

\noindent{\it Case 1}: $i=k$ and $j=l$
\begin{align}
    \bra{ij}\mathcal D^{\rm RWA}\ket{ij} =
    \frac{\sigma}2(V_{ii}+V_{jj}-2V_{ii}V_{jj}).
\end{align}
\noindent{\it Case 2}: $i=j\neq k=l$
\begin{align}
    \bra{ii}\mathcal D^{\rm RWA}\ket{kk} =
    \frac{\sigma}2(-2V_{ik}^2).
\end{align}
\noindent{\it Case 3}: We note that $\omega_i+\omega_{\bar i}=0$ where 
$\bar i = L-i+1$. Therefore, if $l=\bar i$ and $k=\bar j$ the resonance
condition is achieved. To avoid double counting with case 1 we write
$l=\bar i$, $k=\bar j$, $i\neq j$, $i\neq \bar j$ so 
\begin{align}
    \bra{ij}\mathcal D^{\rm RWA}\ket{\bar j\bar i} =
    \frac{\sigma}2(-2V_{i\bar j} V_{j\bar i}) =
    \frac{\sigma}2(-2V_{ij}^2 ).
\end{align}
where we use the fact that $V_{ij}=V_{ji}=V_{\bar i j}=V_{j\bar i}$. 
All the other elements are zero. 

All the non-zero elements of $\mathcal D^{\rm RWA}$ are discusse in Case 1,2,3.
Since most of the terms are zero, it is quite easy to find the eigenvalues of 
$\mathcal D^{\rm RWA}$. We call those eigenvalues $\ket S=\sum_{ij} S_{ij}\ket{ij}$.  
From the cases 1 and 3, one can see that the off-diagonal states where
$S_{ii}=0$ are decoupled from the diagonal ones. Therefore we consider these two
cases separately. Let $\ket{S_o}=\sum_{i\neq j}S_{ij}\ket{ij}$ be an
off-diagonal state, then the eigenvalue equation $\mathcal D^{\rm RWA}\ket{S^o} =
\lambda \ket{S^o}$ written as 
$\bra{kl}\mathcal D^{\rm RWA}\ket{S^o} = \lambda S_{kl}$ for $k\neq l$ is 
\begin{align}
    (V_{kk}+ V_{ll}-2V_{kk}V_{ll}) S_{kl} -2 V_{kl}^2 S_{\bar l\bar k} =
    -\frac2\sigma \lambda_{kl} S_{kl}~,
    \label{e:rwapair}
\end{align}
when $l\neq \bar k$ and 
\begin{align}
    (2V_{kk} -2V_{kk}^2) S_{k\bar k}  =
    -\frac2\sigma \lambda_{k\bar k} S_{k\bar k}~.
\end{align}
Therefore, for each pair $k,l$ Eq \eqref{e:rwapair} is a $2\times2$ matrix
eigenvalue problem whose minimum (in absolute value) eigenvalue is 
\begin{align*}
    \lambda^{\rm min}_{kl}&= -\frac\sigma2(V_{kk}+V_{ll}-4 V_{kk} V_{kl}) \\&
    = -\frac{\sigma}{L+1} \left(
    \sin^2 k_k +\sin^2 k_l - \frac8{L+1}\sin^2 k_k\sin^2 k_l\right).
\end{align*}
On the other hand 
\begin{align*}
    \lambda_{l\bar l}&= -\frac\sigma2(2V_{ll}-2 V_{ll}^2) 
    = -\frac{2\sigma}{L+1} \left(
    \sin^2 k_l - \frac2{L+1}\sin^4 k_l\right).
\end{align*}
When $L\gg 1$ we can neglect the $\mathcal O(L^{-2})$ correction, and since
$V_{ll}$ is minimized for $l=1$  we find that the gap is 
\begin{align}
    {\rm gap}= -\lambda_{1 \bar 1} \approx \frac{2\sigma \pi }{L^3}. 
    \label{e:gapA}
\end{align}

We now show that the other ``diagonal'' eigenvalues $\ket{S^d} = \sum_i
S_{ii}\ket{ii}$ have a larger gap. Writing the eigenvalue equation we find
$
    -\frac{\sigma}2(2 V_{ki} \delta_{ki} S_{kk} -2 V_{ki}^2 S_{ii}) = \lambda
    S_{kk}
    $, 
    namely we have to find the eigenvalues of the matrix $R_{ik}= 
    \sigma ( V_{ik}\delta_{ik} - V_{ik}^2)$. Calling $V^d= \sigma {\rm diag} V$
    and $a_i = \frac{2\sqrt{\sigma}}{L+1}\sin^2 k_i$ then $R=-V^d+ a^T a$. Using
    the matrix determinant lemma in the eigenvalue equation we find
    \begin{align}
      \nonumber
        0 &= \det(\lambda \openone + V^d - a^T a) \\&= \det(\lambda\openone+V^d)
        \left(1-a^T \frac{1}{\lambda\openone+V^d}a\right)~.
        \label{e:equationgap}
    \end{align}
    The first term in the above equation gives the solutions
    $\lambda=-V_{ll}=-\frac{2\sigma}{L+1}\sin^2 k_l$ which have a higher gap. 
    On the other hand, the second term in \eqref{e:equationgap} provides the
    equation
    \begin{align*}
        0 
        &= 1- \frac{4\sigma}{(L+1)^2}\sum_l \frac{\sin^4 k_l}{\lambda+
            \frac{2\sigma}{L+1}\sin^2 k_l} 
            \\
            &= 1- \frac{2}{(L+1)}\sum_l \frac{\sin^4 k_l}{\frac{L+1}{2\sigma}\lambda+
            \sin^2 k_l} 
            \\
            &= 1- \frac{2}{(L+1)}\sum_l \left(
            \sin^2 k_l+ \frac{L+1}{2\sigma}\lambda
            \frac{\sin^2 k_l}{\frac{L+1}{2\sigma}\lambda + \sin^2 k_l} 
            \right)~,
    \end{align*}
    where in the last equation we use the identity $\frac{1}{a+b}=\frac1a-
    \frac{b}{a(a+b)}$. Since $\sum_l \sin^2 k_l = (L+1)/2$ we are left with the
    equation 
    \begin{align}
        0 = \lambda \sum_l 
            \frac{\sin^2 k_l}{\frac{L+1}{2\sigma}\lambda + \sin^2 k_l} .
    \end{align}
    A solution to that equation is clearly $\lambda=0$, namely the steady state. 
    On the other hand all the other solutions must satisfy $\lambda < - 
    \frac{2\sigma}{L+1}\sin^2 k_l$ for some $l$ because otherwise all the elements in the sum
    are positive and there is clearly no solution. Therefore all the solutions
    must satisfy $|\lambda|> \frac{2\sigma}{L+1}\sin^2 k_1 > {\rm gap}$.
    This concludes the proof that the gap is given by \eqref{e:gapA}.

\section{Strong driving limit}
\label{s:strong}

We focus here in the derivation of the effective Liouvillean \eqref{e:Lstrongsym}. 
Let us define then $\mathcal P$ as the
projector onto the low-energy (eigenvalue zero) subspace of 
$\mathcal D=\frac12 
(n^\uparrow_1 - n^\downarrow_1) (n^\uparrow_1 - n^\downarrow_1)$.
This space is generated by all the
states such that $n_1^\uparrow=n_1^{\downarrow}$. We set also $\mathcal Q=
\openone-\mathcal P$ and call $\mathcal H$ the Hamiltonian part such that 
$\mathcal L_q= -i \mathcal H - \sigma\mathcal D$. We call then also $X_{PP} =
\mathcal P X\mathcal P$, with similar definitions for 
$X_{PQ}$, $X_{QP}$, $X_{QQ}$. We can therfore write $\mathcal L_q$ in the
block form
\begin{align}
  \mathcal L_q = \begin{pmatrix} 
    -\ri \mathcal H_{PP} & 
    -\ri \mathcal H_{PQ} \\ 
    -\ri \mathcal H_{QP} & 
    -\ri \mathcal H_{QQ} - \sigma \mathcal D_{QQ} 
  \end{pmatrix}~,
  \label{e:LdecPQ}
\end{align}
where $\sigma\gg\|\mathcal H\| ,\|\mathcal D\|$ and where we used the fact
that $\mathcal P\mathcal D=\mathcal D\mathcal P =0$. The low-energy eigenvalues
can then be obtained using the determinant identity $\det{\begin{pmatrix}
    A&B\\C&D\end{pmatrix}} = \det(D)\det(A-BD^{-1}C)$ 
-- see also \cite{zanardi2016dissipative,zanardi2014coherent} for a related
approach. Indeed, using a
first order expansion for $\sigma\to\infty$ it is simple to see that the
small eigenvalues are the eigenvalues of the effective
operator 
\begin{align}
    \mathcal L_q^{\rm eff.} = -i \mathcal H_{PP} - \frac1\sigma 
    \mathcal H_{PQ} \mathcal D_{QQ}^{-1} \mathcal H_{QP}   .
    \label{e:leff}
\end{align}
The above effective operator can be obtained also with a 
(possibly non-unitary) similarity transformation $e^{\mathcal S_{\rm D}}$ 
to decouple the ``low-energy'' and 
``high-energy'' subspaces. Namely one can find $\mathcal S_{\rm D}$ such that
\begin{align}
  \begin{pmatrix}
    \mathcal L_q^{\rm eff} & 0 \\  0 & \mathcal O(\sigma)  
  \end{pmatrix} &= 
  e^{\mathcal S_{\rm D}} \mathcal L_q e^{-\mathcal S_{\rm D}} 
  \label{e:simill}
  \\ \nonumber& =
  \mathcal L_q + [\mathcal S_{\rm D}, \mathcal L_q] + \frac{
  [\mathcal S_{\rm D}, [\mathcal S_{\rm D}, \mathcal L_q]]}2  +
  \mathcal O(\|\mathcal S_{\rm D}\|^3)~.
\end{align}
One finds that \eqref{e:simill} is valid 
up to the first order in $\sigma^{-1}$, with $\mathcal L_q^{\rm eff}$
given by \eqref{e:leff}, by choosing 
\begin{align}
  \mathcal S_{\rm D}&=\frac{\mathcal S_1}\sigma +
\frac{\mathcal S_2}{\sigma^2} + \mathcal O(\sigma^{-3})~,
  \label{e:Sleff}
\end{align}
such that 
\begin{align*}
  \mathcal S_1 &= \begin{pmatrix}
    0 & i\mathcal H_{PQ} \mathcal D_{QQ}^{-1}
    \\ -i\mathcal D_{QQ}^{-1} \mathcal H_{QP} &0 
  \end{pmatrix}~,
  &
  \mathcal S_2 &= \begin{pmatrix}
    0 & \mathcal S_{2,*}
    \\ -\mathcal S_{2,*}^{\dagger} & 0 
  \end{pmatrix}~,
\end{align*}
where $ \mathcal S_{2,*}= \mathcal H_{PP}\mathcal H_{PQ}\mathcal D_{QQ}^{-2}
-\mathcal H_{PQ}\mathcal D_{QQ}^{-1}\mathcal H_{QQ}\mathcal D_{QQ}^{-1}$.  
Note that $i\mathcal S_1$ is a Hermitian operator, unlike
$i\mathcal S_2$.

We now obtain the effective operator explicitly. 
Since $\mathcal P$ commutes with all the operators acting on all but the first
sites, one realizes that $\mathcal H_{PQ}$ and $\mathcal H_{PQ}$  are only
composed by the projections of $a_{1\updownarrow}^{\dagger}a_{2\updownarrow}$
and their complex conjugate. Moreover,
\begin{align*}
    \mathcal P a_{1\uparrow}^{\dagger} \mathcal Q &= 
    \sum_{n_{1\uparrow}} 
    \sum_{m_{1\uparrow}\neq m_{1\downarrow}} \ket{n_{1\uparrow}n_{1\uparrow}}
    \bra{n_{1\uparrow}n_{1\uparrow}} a_{1\uparrow}^{\dagger}
    \ket{m_{1\uparrow}m_{1\downarrow}}
    \bra{m_{1\uparrow}m_{1\downarrow}}
    \\ &= 
    \sum_{n_1} \sqrt{n_1} 
    \ket{n_{1},n_{1}}
    \bra{n_{1}{-}1,n_{1}} ~,
\end{align*}
where the $\ket{mn}$ is a short-hand notation for $\frac{
(a_{1\uparrow}^\dagger)^m
(a_{1\downarrow}^\dagger)^n}{\sqrt{m!n!}}\ket 0$. Similarly we find 
\begin{align}
    \mathcal P a_{1\downarrow}^{\dagger} \mathcal Q &= 
    \sum_{n_1} \sqrt{n_1} \ket{n_{1},n_{1}} \bra{n_{1},n_{1}{-}1} ~,
    \\
    \mathcal P a_{1\uparrow} \mathcal Q &= 
    \sum_{n_1} \sqrt{n_1} \ket{n_{1},n_{1}} \bra{n_{1}{+}1,n_{1}} ~,
    \\
    \mathcal P a_{1\downarrow} \mathcal Q &= 
    \sum_{n_1} \sqrt{n_1+1} \ket{n_{1},n_{1}} \bra{n_{1},n_{1}{+}1} ~.
\end{align}
Since in $\mathcal H_{QP}$ the up/down states on the first site differ only for
one paritcle it is $\mathcal D_{QQ}^{-1} \mathcal H_{QP} = 2 \mathcal H_{QP}$.
Hence the effective operator is given by $
 -i \mathcal H_{PP} - \frac2\sigma 
    \mathcal H_{PQ} \mathcal H_{QP}
$. This can be computed from 
\begin{align}
    \mathcal P a_{1\updownarrow}^\dagger \mathcal Q a_{1\updownarrow} \mathcal P
    &= n_{1\updownarrow} \mathcal P \\
    \mathcal P a_{1\updownarrow} \mathcal Q a_{1\updownarrow}^\dagger \mathcal P
    &= (n_{1\updownarrow}+1) \mathcal P \\
    \mathcal P a_{1\uparrow}^\dagger \mathcal Q a_{1\downarrow}^\dagger \mathcal P &=
    a_{1\uparrow}^\dagger  a_{1\downarrow}^\dagger \mathcal P. 
\end{align}
and their Hermitian conjugate (all the other terms are zero). Moreover, 
$n_{1\uparrow}\mathcal P = n_{1\downarrow}\mathcal P $.  We find then
\begin{align}
    \mathcal H_{PQ} \mathcal H_{QP}  =&   -2(
        a_{1\uparrow}^\dagger a_{1\downarrow}^\dagger
        a_{2\uparrow} a_{2\downarrow} + {\rm h.c.})  + 
        n_{1\uparrow}(n_{2\downarrow} + 1)
        \\ &  + 
        n_{1\downarrow}(n_{2\uparrow} + 1)+
        n_{2\downarrow}(n_{1\uparrow} + 1)+
        n_{2\uparrow}(n_{1\downarrow} + 1)
     \nonumber 
    \\ \nonumber  
    = & -2(
        a_{1\uparrow}^\dagger a_{1\downarrow}^\dagger
        a_{2\uparrow} a_{2\downarrow} + {\rm h.c.}) -1 + 
        \\ & + 
        (n_{1\uparrow} + n_{1\downarrow} + 1)
        (n_{2\uparrow} + n_{2\downarrow} + 1) 
\label{e:openbeforerwa}
\end{align}

In order to make further analytical progresses we also use the rotating wave
approximation 
which is consistent with the perturbative treatment (see Appendix \ref{s:weak}) 
since 
$\mathcal L_q^{\rm eff} =  -i \mathcal H_{PP} - \frac2\sigma 
    \mathcal H_{PQ} \mathcal H_{QP}$ and $2/\sigma$ is small. 
We note that 
$ \mathcal H_{PP} = \sum_{\alpha=2}^L (a_{\alpha\uparrow}^\dagger a_{\alpha{+}1,\uparrow} 
    - a_{\alpha\downarrow}^\dagger a_{\alpha{+}1,\downarrow} + {\rm h.c.})$. 
The above operator can be diagonalized with a Bogoliobov transformation:
defining the operators $\tilde a_{k \updownarrow} = \sum_{\alpha=1}^{L-1}
\frac2L \sin^2\left(\frac{\pi k\alpha}L\right) a_{\alpha{+}1,\updownarrow}$ we
find that 
$\mathcal H_{PP} = \sum_{k=1}^L 2\cos\left(\frac{k\pi}{L}\right) (\tilde
n_{k\uparrow}-\tilde n_{k\downarrow})$. Because of this particular form, the
rotating wave approximation in \eqref{e:openbeforerwa} corresponds to expanding 
the operators $a_{2\updownarrow}$ into the diagonal basis $\tilde
a_{k\updownarrow}$, neglecting the ``oscillating'' off-diagonal terms. In other 
terms, we can write 
\begin{align}
  \mathcal L^{\rm eff}_q = \hat{\mathcal L}_q + \mathcal L_q^{\rm osc.}~,
\end{align}
where $  \hat{\mathcal L}_q$ is the Hermitian Liouvillean in the rotating 
wave approximation 
shown in Eq.~\eqref{e:Lstrongsym}, where
$\mathcal O(\hat{\mathcal L}_q) =\mathcal O(\sigma^{-1})$, 
while $\mathcal L_q^{\rm osc.}$, of order $\mathcal O(\sigma^{0})$, is composed 
by the oscillating terms that are neglected in the long-time limit. 
In particular, from \eqref{e:RWAlarget} one finds that RWA
holds for $t\gg\mathcal O(L^2)$. This approximation is therefore consistent with 
the results of section \ref{s:chain}, where one finds a Liouvillean gap 
$\mathcal O(L^{-3})$ that provides a lower bound to the convergence time 
$t>\mathcal O(L^3)$. 
However, while the eigenvalues depend only on the Hermitian operator $\hat{\mathcal L}_q$,
the eigenvectors depend on the oscillating terms via 
\eqref{e:Srwa}. By mixing \eqref{e:RWAev} with \eqref{e:simill} we find then
that the eigenvalues with small $\mathcal O(\sigma^{-1})$ real part have
right eigenvectors given by
\begin{align}
  e^{\mathcal S} \ket{\Phi_i}~,
  \label{e:evstrong}
\end{align}
where $\ket{\Phi_i}$ form an orthonormal basis (depdent on both $H$ and $V$), 
$\mathcal S \approx \mathcal S_{\rm RWA} + \mathcal S_{\rm D} + 
[ \mathcal S_{\rm RWA},  \mathcal S_{\rm D}]/2 
= \mathcal O(\sigma^{-1})$, 
but $e^{\mathcal S}{}^\dagger\neq e^{-\mathcal S}$. 
The corresponding left eigenvectors are then 
$\bra{\Phi_i}e^{-\mathcal S}$.

\section{Diagonalization of the Richardson-Gaudin model}
\label{s:richardson}
We perform explicitly the diagonalization of the 
Richardson-Gaudin model \eqref{e:gaudin} in the bosonic
representation discussed in Section~\ref{s:symm}, where
$K_i^- = \tilde a_{i\uparrow}
\tilde a_{i\downarrow}$, $K_i^+ = (K_i^-)^\dagger$ and 
$K_i^z = (\tilde n_{i\uparrow} + \tilde n_{i\downarrow} + 1)/2$.
We start by defining a trial eigenstate $\ket{\Omega_\nu}$ with no pairing,
namely such that 
\begin{align}
    K_i^-\ket{\Omega_\nu}  &= 0~, &  
    K_i^z\ket{\Omega_\nu}  &= \nu_i \ket{\Omega_\nu} ~. 
\end{align}
These equations force the constraints
\begin{align}
    \nu_i & = (n_{i\uparrow} + n_{i\downarrow} +1)/2~, &
    n_{i\uparrow}n_{i\downarrow}&=0~,
    \label{e:gaudinconstraint1}
\end{align}
namely there cannot be in the same site both up-particles and down-particles.
Moreover, $\nu_0\equiv 1/2$ because the model has been obtained by projecting
the Liouvillean into the states where $n_{0\uparrow} = n_{0\downarrow}$. The
eigenvalue of state $\ket{\Omega_\nu}$  is thus 
\begin{align}
    \hat{\mathcal L_q} \ket{\Omega_\nu} &= E_0 \ket{\Omega_\nu}~,
    \\E_0 &=\frac2\sigma - \frac 8\sigma  \sum_k g_k \nu_0 \nu_k 
   =\frac2\sigma - \frac 4\sigma  \sum_k g_k \nu_k ~.
\end{align}
Since there are extra constraints, $q= \sum_{i=0}^{L-1} n_{i\uparrow} = 
\sum_{i=0}^{L-1} n_{i\downarrow}$, for a given set of allowed ``quantum numbers''
$v_k$ the number $N$ of paired particles satisfies $\sum_i (2v_i-1) + 2 N =2q$,
namely
\begin{equation}
    N = q-\sum_i \left(v_i-\frac12\right). 
    \label{e:paired}
\end{equation}
By defining the {\it ansatz}
\begin{align}
    \ket\psi &= \prod_{\alpha=1}^N C^+_\alpha \ket{\Omega_\nu} & 
    & C^+_\alpha = \sum_{j=0}^{L} u_{j\alpha} K_j^+~.
    \label{e:prodC}
\end{align}
one sees that 
\begin{align}
    \hat{\mathcal L_q} \ket \psi \nonumber
    &= E_0\ket\psi + [\hat{\mathcal L_q}, \prod_\alpha C^+_\alpha] \ket{\Omega_\nu}
    \\
&= E_0\ket\psi + \sum_\alpha\left( \prod_{\gamma\neq\alpha} C^+_\gamma\right) 
[\hat{\mathcal L_q}, C^+_\alpha] \ket{\Omega_\nu}
\\ &\quad\quad + \frac12 \sum_{\alpha\neq\beta} \left( \prod_{\gamma\neq\alpha,\beta} C^+_\gamma\right) 
[[\hat{\mathcal L_q}, C^+_\alpha], C^+_\beta] \ket{\Omega_\nu}~.
\nonumber
\end{align}
Moreover, 
\begin{align}
  \nonumber
[\hat{\mathcal L_q}, C^+_\alpha]  &= -\frac8\sigma \sum_k g_k
(u_{0\alpha}-u_{k\alpha})(K_0^+K_k^z-K_0^zK_k^+)~,
\\
[[\hat{\mathcal L_q}, C^+_\alpha], C_\beta^+]  &= \frac8\sigma \sum_k g_k
(u_{0\alpha}-u_{k\alpha})
(u_{0\beta}-u_{k\beta})K_0^+K_k^+~.
\label{e:commC2}
\end{align}
We now first consider the $N=1$ case and impose the eigenvalue equation 
$\hat{\mathcal L_q}\ket\psi=\lambda\ket\psi$ where we define $\lambda = E_0 -
\frac 8\sigma \sum_\alpha E_\alpha$. The eigenvalue equation becomes then 
\begin{equation*} 
    \sum_k g_k (u_{0\alpha}-u_{k\alpha})(K_0^+\nu_k-\nu_0K_k^+) = 
    E_\alpha (u_{0\alpha} K_0^+ + \sum_k u_{k\alpha} K_k^+)~.
\end{equation*}
From that equation we get the relationship 
\begin{equation}
    -\nu_0 g_k (u_{0\alpha}-u_{k\alpha}) = u_{k\alpha} E_\alpha,
\end{equation}
namely 
\begin{align}
    u_{k\alpha} &= \frac{\nu_0 g_k u_{0\alpha}}{\nu_0 g_k - E_\alpha},\\
    u_{0\alpha}- u_{k\alpha} &= -\frac{E_\alpha u_{0\alpha}}{\nu_0 g_k -
    E_\alpha} = -\frac{E_\alpha}{\nu_0 g_k} u_{k\alpha}. 
\end{align}
By using the last equation we find 
\begin{align*}
[[\hat{\mathcal L_q}, C^+_\alpha], C_\beta^+]  &= \frac8\sigma \sum_k g_k
\frac{E_\alpha u_{0\alpha}}{\nu_0 g_k -
    E_\alpha}
    \frac{E_\beta u_{0\beta}}{\nu_0 g_k -
    E_\beta} K_0^+K_k^+
    \\ &=  \frac8{\sigma\nu_0} \frac{E_\alpha E_\beta}{E_\alpha - E_\beta} \sum_k (
    u_{0\beta}u_{k\alpha}- 
    u_{k\beta}u_{0\alpha})
    K_0^+K_k^+
    \\ &=  \frac8{\sigma\nu_0} \frac{E_\alpha E_\beta}{E_\alpha - E_\beta} (
    u_{0\beta}C_{\alpha}^+- 
    C_{\beta}^+ u_{0\alpha})
    K_0^+
     \\&=  \frac{8}{\sigma} \left(M_{\alpha\beta} K_0^+ C_\alpha^+ + 
     M_{\beta\alpha} K_0^+ C_\beta^+\right)
    ~.
\end{align*}
where $M_{\alpha\beta}= \frac{E_\alpha E_\beta}{E_\alpha - E_\beta}
\frac{u_{0\beta}}{\nu_0}$. Using all the above results the eigenvalue equation
becomes
\begin{align}
    (\hat{\mathcal L_q}-\lambda)\ket{\psi}=& -\frac8{\sigma} 
    \sum_\alpha\left( \prod_{\gamma\neq\alpha} C^+_\gamma\right) Z_\alpha
    \ket{\Omega_\nu}~,
    \\
    Z_\alpha =& \sum_k \frac{g_k E_\alpha u_{0\alpha}}{\nu_0 g_k -E_\alpha}
    ( \nu_0K_k^+ - K_0^+\nu_k)
    \\ &  -\sum_{\beta\neq\alpha} M_{\beta\alpha} K_0^+ - E_\alpha \left(
    u_{0\alpha} K_0^+ + 
    \sum_k u_{k\alpha} K_k^+\right) \nonumber .
\end{align}
By
evaluating $Z_\alpha=0$ one gets the equations 
\begin{align}
    \sum_k \frac{g_k \nu_k E_\alpha}{E_\alpha -\nu_0 g_k} + \frac1{\nu_0}
    \sum_{\beta\neq\alpha}\frac{E_\alpha E_\beta}{E_\alpha - E_\beta}
= E_\alpha~,
\label{e:gaudineq}
\end{align}
for $\alpha=1,\dots,N$ where $N$ is given by \eqref{e:paired}. Clearly,
$E_\alpha=0$ is a solution, while the solutions different from zero are found by
solving the equation 
\begin{align}
    \sum_k \frac{2 g_k \nu_k }{2E_\alpha -g_k} + 2
    \sum_{\beta\neq\alpha}\frac{E_\beta}{E_\alpha - E_\beta}
= 1~,
\label{e:gaudineqNZ}
\end{align}
where we used the fact that $\nu_0=1/2$. 
In conclusion, the eigenvalues of the
Liouvillean $\hat{\mathcal L_q}$ are 
\begin{align*}
    \lambda = E_0 - \frac8\sigma \sum_\alpha E_\alpha 
    = -\frac{2}\sigma \left(\sum_k g_k n_k   +
    4\sum_{\alpha=1}^{q-\sum_k n_k/2} E_\alpha\right)
 ~,
\end{align*}
where $n_k = 2\nu_k-1$ and the $E_\alpha$ are either zero or the solution of 
\eqref{e:gaudineqNZ}. From that expression it is clear that the steady state
corresponds to $E_\alpha=0$ and $n_k=0$. The eigenvalues for larger values of
$q$ are given by all the previous solutions with smaller $q$ (this can be seen 
by adding some $E_\beta=0$ for larger values of $N$) together with new solutions
due to the larger values of $N$ and the larger set of allowed configurations for
$n_k$.

\section{Solution of the SU(2q)-invariant Gaudin model}
\label{s:gaudinsuq} 

We describe here the  algebraic approach to general Gaudin models and then
apply it to our  general fermionic representation introduced in
Section~\ref{s:gapgen}. 
We fix a basis $h_\alpha^{(j)}$ of the Cartan subalgebra acting on the $j$-th copy formed by the diagonal operators $X^{(j)}_{\alpha\alpha}$. 
A state $\ket{\Omega_\nu}$  which is a simultaneous eigenvector 
of all the operators $h_\alpha^{(j)}$ is called a weight vector. We write
$h^{(j)}_\alpha \ket{\Omega_\nu} = \nu_{j\alpha} \ket{\Omega_\nu}$ where $\nu^j_{\alpha}$ is 
called {\it weight}. 
On the other hand, in the Cartan-Weyl basis 
the eigenvalue $\chi$ of the adjoint transformation, namely
$[h^{(j)}_\alpha, e_\chi] = \chi_{j\alpha} e_\chi$, for a given $e_\chi$ in the representation,
is called a {\it root}. Because of Eq.\eqref{e:sualgebra} a root can only have eigenvalue -1,0,1.  
If one fixes an ordering $c_\alpha>c_{\alpha+1}$ and writes
$h^{(j)}=\sum_\alpha c_\alpha h^{(j)}_\alpha$
then the eigen-operators of $h^{(j)}$ with positive eigenvalue are called the 
``raising operators''. They correspond to 
$X^{(j)}_{\alpha\beta}$ for any $\alpha<\beta$. A highest weight vector is a weight
vector $\ket{\Omega_\nu}$ such that all the other vectors in an irreducible representation can 
be obtained from $\ket{\Omega_\nu}$ via some lowering operators.  As such, a highest weight
vector is annihilated by all the raising operators. 
We call $\chi^j$ the simple roots of the algebra, and we fix an inner product between
roots $(\chi^j,\chi^k)=\sum_\alpha \chi^j_\alpha\chi^k_\alpha$, and write 
$|\chi^j|^2 = (\chi^j,\chi^j)$. The matrix $C_{jk} = 
\frac{2(\chi^j,\chi^k)}{|\chi^j|^2}$ is called the Cartan matrix. We 
call also $F_{jk}=(C^{-1})_{jk} \frac{2}{|\chi^k|^2 }$.  
Moreover, we call $z_0 = 0$ and $z_k=2 g_k^{-1}$ for $k>0$. 

Thanks to the above definitions, and owing to the results of Refs.~\cite{ushveridze1994quasi,falceto1997unitarity}, 
we can write the eigenvalues of the Gaudin model \eqref{e:gaudingl} as 
\begin{align}
\lambda_{\{\mu\} }   = -\frac{2q}\sigma+\frac8\sigma\Big[&
\sum_{k=1}^{L-1}  \frac{\sum_{ij} \mu^0_i F_{ij} \mu^k_j}{z_k-z_0}+
\label{e:eiggaudin}
\\ \nonumber & 
+ \sum_{j=1}^{2q-1}\sum_\alpha \frac{|\chi^j|^2}2\frac{\mu^0_j}{z_0-\omega_{j\alpha} }
\Big],
\end{align}
where $\mu^j_k$ are the eigenvalues of the Chevalley operators $H^{(j)}_k = 
\frac{2}{|\chi^k|^2}\sum_\alpha \chi^k_\alpha h^{(j)}_{\alpha}$, namely
$H^{(j)}_k \ket{\Omega_\nu} = \mu^j_k \ket{\Omega_\nu}$ and so $\mu^j_k = 
\frac{2}{|\chi^k|^2}\sum_\alpha \chi^k_\alpha \nu^j_{k}$, and 
where the Bethe roots satisfy the equations
\begin{align}
\sum_{i\beta} \frac{C_{ij}}{\omega_{i\beta}-\omega_{j\alpha}} = \sum_{k=0}^{L-1} 
\frac{\mu^k_j}{z_k- \omega_{j\alpha}}~.
\label{e:bethegen}
\end{align}

The above expressions for the eigenvalues hold whenever the operators $X^{(j)}_{\alpha\beta}$ define  any semi-simple Lie algebra. In the particular case discussed in Section~\ref{s:gapgen} those operators define a SU(2q)-invariant Gaudin model, in a specific multi-fermion representation. 
For SU(2q) the simple roots are $\chi^j_\alpha = \delta_{\alpha j}-\delta_{\alpha+1,j}$ so 
$|\chi^j|^2=2$ and  
$C_{ij}= 2 \delta_{ij}-(\delta_{i,j-1}+ \delta_{i,j+1})$, where $i,j=1,\dots 2q{-}1$. 
Therefore, $F_{ij} = \sum_\ell \frac2{2q}
\frac{\sin(\pi i\ell/ 2q)\sin(\pi j\ell/2q)}{2-2 \cos(\ell \pi/2q)}$ and the 
Chevalley operators are given by 
$H^{(j)}_\alpha = X^{(j)}_{\alpha,\alpha}- X^{(j)}_{\alpha+1,\alpha+1}$. 
We fix the ordering $\{({\downarrow},1), ({\downarrow},2)\dots, ({\uparrow},1), ({\uparrow},2), \dots\}$
so that 
\begin{align*}
H^{j}_\alpha = \begin{cases}
-\tilde{a}^\dagger_{j,\alpha,\downarrow} \tilde{a}_{j,\alpha,\downarrow} +
\tilde{a}^\dagger_{j,\alpha+1,\downarrow} \tilde{a}_{j,\alpha+1,\downarrow} 
& \text{for  }  1\le\alpha\le q-1~,\\
\tilde{a}^\dagger_{j,\alpha,\uparrow} \tilde{a}_{j,\alpha,\uparrow} -
\tilde{a}^\dagger_{j,\alpha+1,\uparrow} \tilde{a}_{j,\alpha+1,\uparrow}
& \text{for  }  1\le\alpha-q\le q-1~,\\
1- 
\tilde{a}^\dagger_{j,q,\downarrow} \tilde{a}_{j,q,\downarrow} -
\tilde{a}^\dagger_{j,1,\uparrow} \tilde{a}_{j,1,\uparrow} 
& \text{for  }  \alpha=q~.
\end{cases}
\end{align*}
Because of the above equations, 
the raising operators  are given by $a^\dagger_{j,i,\uparrow} a_{j,k,\uparrow}$ with 
$i>k$, by  $a^\dagger_{j,i,\downarrow} a_{j,k,\downarrow}$ with 
$i<k$, and by $a_{j,i,\sigma} a_{j,k,\sigma}$. Therefore, the highest weight vectors 
may contain in the same mode $j$ either spin-$\uparrow$ particles or  spin-$\downarrow$ particles, but not both. The only possible highest weight states  
are then either $\prod_i^{n_{j\uparrow}} a^\dagger_{j,i,\uparrow}\ket{0}$ or
$\prod_i^{n_{j\downarrow}} a^\dagger_{j,q-i+1,\downarrow}\ket{0}$. 
These states are parametrized by 
the numbers $n_{j\uparrow}$ and $n_{j\downarrow}$ that satisfy 
$n_{j\uparrow}n_{j\downarrow} =0$. Therefore
\begin{align}
\mu^k_j &= \delta_{j,q}(1-\delta_{n_{\downarrow k}>0} - \delta_{n_{\uparrow k}>0})
+ \delta_{j,q+n_{\uparrow k}} + \delta_{j,q-n_{\downarrow k}}~.
\end{align}
Moreover, $n_{0\uparrow}=n_{0\downarrow}$ so 
$ \mu^0_j = \delta_{j,q} $. 
By explicit calculation for $j\le q$ one finds $F_{qj} = F_{q,2q-j} = j/2$.
Therefore, \eqref{e:eiggaudin} becomes 
\begin{align}
\lambda_{\{n\} }   &= -\frac{2q}\sigma+\frac8\sigma\Big[
\sum_{k=1}^{L-1}  \frac{g_k}2\Big(\frac{q}2
(1-\delta_{n_{\downarrow k}>0} - \delta_{n_{\uparrow k}>0})    +
\nonumber \\&\hspace{2cm}
\frac{(q-n_{\downarrow k})\delta_{n_{\downarrow k}>0}  
	+ (q-n_{\uparrow k})\delta_{n_{\uparrow k}>0} }2\Big)
\nonumber \\&\hspace{2cm}
- \sum_\alpha \frac{1}{\omega_{q,\alpha} }
\Big] 
\\
&= -\frac2\sigma\left[
\sum_{k=1}^{L-1}  g_k\left(
n_{\downarrow k} + n_{\uparrow k}   \right)
+ 4 \sum_\alpha \frac{1}{\omega_{q,\alpha} } \right]~,
\end{align}
where 
 $\omega_{j,\alpha}$ are the solutions of Eq.~\eqref{e:bethegen}, namely of Eq.~
 \eqref{e:bethesuq}.

\section{Explicit mean-field analysis}
\label{s:emf}

In this section we perform explicitly the mean-field calculations 
discussed in section \ref{s:mf}, and we closely follow the notation of that section. 
We remind that Eq.~\eqref{e:lsb} can be written as 
\begin{align}
    \mathcal L_q & = \sum_{i} \lambda_{i}\; \tilde a'_i \tilde a_i
    - \frac\sigma2\sum_{i,j,k,l} \tilde V_{ij}\tilde V_{kl}
    \;
    \tilde a'_i\tilde a'_k \tilde a_j \tilde a_l~, 
    \label{e:lsbbasis}
\end{align}
where the $\lambda$'s are ordered with decreasing (negative) real part, 
$\lambda_0=0$, 
$\mathring V = Z\tilde V Z^{-1}$ and we remind that the new bosonic 
creation operators are obtained via the non-unitary Bogoliubov 
transformation 
$ \tilde{a}'_i = \sum_\alpha Z_{\alpha i} a_\alpha^\dagger$,  
$ \tilde{a}_i = \sum_\alpha (Z^{-1})_{i\alpha} a_\alpha$. 
The steady state is therefore the 
boson condensate $\ket{\Omega}=\frac{(\tilde a_0')^q}{\sqrt q!}\ket 0$ where
$\ket 0$ is the bosonic vacuum. Indeed, clearly this state is annihilated by the
quadratic term. To see that even the second one annihilates it is important to
remind that $S^{-1}_{0\alpha}$ is the right eigenvector of the steady state
(corresponding to the steady state) and the corresponding left eigenvalue 
$S_{\alpha 0}$ is the identity operator. Therefore, 
$\tilde  V_{i0} = \sum_{\alpha\beta} 
S^{-1}_{j \alpha} \mathring V_{\alpha\beta} S_{\beta 0} = 0$ since 
$\sum_\beta \mathring V_{\alpha\beta} S_{\beta 0}$ is a vectorization of the
expression $[V,\openone]$. Similarly 
$\tilde V_{0i}=0$. 
To study the elementary excitations with respect
to this state, one can use Bogoliubov (mean field) approach 
starting from the variational states 
$\ket{\psi} =  \sum_j \psi_j \frac{(\tilde a_0')^{q-1}}{\sqrt{(q-1)!}}a'_j\ket 0$, for $j\neq 0$
and the corresponding 
$\bra{\psi'} =  \sum_j \psi'_j \bra0\frac{(\tilde a_0)^{q-1}}{\sqrt{(q-1)!}}a_j$, 
where $\sum_j \psi'_j \psi_j = 1$. 
The variational Liouvillean then becomes 
\begin{align*}
  \mathcal L^V &= 
  \bra{\psi'}\mathcal L_q\ket{\psi} \\&= \sum_j \lambda_j \psi'_j\psi_j 
      - \frac\sigma2\sum_{i,j,k,l} \tilde V_{ij}\tilde V_{kl}
    \;
  \bra{\psi'}
    \tilde a'_i\tilde a'_k \tilde a_j \tilde a_l
  \ket{\psi} ~,
\end{align*}
which, similarly to the Rayleigh-Ritz method, 
has to satisfy $\partial_\psi \mathcal L^V = \partial_{\psi'} \mathcal L^V=0$ 
with the constraint $\sum_j \psi_j'\psi_j=1$ (see e.g. \cite{laestadius2017analysis}). 
However, because 
$\tilde{V}_{i0}=\tilde{V}_{0i}=0$ for all $i$, one can restrict the sum in the above equation
to the values $i,j,k,l>0$, but because there is only one particle in $\ket{\psi}$ 
in the states $i>0$ one finds that 
\begin{align*}
  \mathcal L^V &= \sum_j \lambda_j \psi'_j\psi_j ~,
\end{align*}
namely that in the single-excitation subspace the variational Liouvillean is already diagonal. 
This shows that the eigenvalues, at least in the low-energy subspace, are not ``renormalized''
for larger values of $q$.

\end{document}